\documentclass[useAMS,usenatbib]{mn2e}
\usepackage{graphicx}
\usepackage{times}
\usepackage{mn2e-breakabs}
\usepackage{amssymb}
\usepackage{amsmath}
\usepackage{url}

\title[Clustering in LOWZ BOSS galaxies]{The clustering of galaxies in the SDSS-III Baryon Oscillation Spectroscopic Survey: galaxy clustering measurements in the low redshift sample of Data Release 11}

\author[Tojeiro et al.]{
 \parbox{\textwidth}{Rita Tojeiro$^1$\thanks{E-mail: rita.tojeiro@port.ac.uk},
Ashley J. Ross$^1$,
Angela Burden$^1$,
Lado Samushia$^1$,
Marc Manera$^2$,
Will J. Percival$^1$,
Florian Beutler,$^3$
Antonio J. Cuesta$^{4,5}$,
Kyle Dawson$^6$,\\
Daniel J. Eisenstein$^7$,
Shirley Ho$^8$,
Cullan Howett$^1$,
Cameron K. McBride$^7$,\\
Francisco Montesano$^9$,
John K. Parejko$^4$,
Beth Reid$^{3,10,11}$,
Ariel G. S\'anchez$^9$,\\
David J. Schlegel$^3$,
Donald P. Schneider$^{12,13}$,
Jeremy L. Tinker$^{14}$,
Mariana Vargas Maga\~na$^{8}$,
Martin White$^{3,10,15}$
}\vspace{3mm}\\
$^1$Institute of Cosmology and Gravitation, Dennis Sciama Building, University of Portsmouth, Burnaby Road, Portsmouth, PO1 3FX, UK \\
$^2$University College London, Gower Street, London WC1E 6BT, UK\\
$^3$Lawrence Berkeley National Laboratory, 1 Cyclotron Road, Berkeley, CA 94720, USA\\
$^4$Department of Physics, Yale University, 260 Whitney Ave, New Haven, CT 06520, USA\\
$^5$Institut de Ci\`encies del Cosmos, Universitat de Barcelona, IEEC-UB, Marti i Franqu\`es 1, E08028 Barcelona, Spain\\
$^6$Department Physics and Astronomy, University of Utah, UT 84112, USA\\
$^7$Harvard-Smithsonian Center for Astrophysics, 60 Garden St., Cambridge, MA 02138, USA\\
$^8$Department of Physics, Carnegie Mellon University, 5000 Forbes Avenue, Pittsburgh, PA 15213, USA\\
$^9$Max-Planck-Institut  f\"uer extraterrestrische Physik, Postfach 1312, Giessenbachstr., 85748 Garching, Germany\\
$^{10}$Hubble fellow\\
$^{11}$Department of Physics, University of California, 366 LeConte Hall, Berkeley, CA 94720, USA\\
$^{12}$Department of Astronomy and Astrophysics, The Pennsylvania State University, University Park, PA 16802, USA\\
$^{13}$Institute for Gravitation and the Cosmos, The Pennsylvania State University, University Park, PA 16802, USA\\
$^{14}$Center for Cosmology and Particle Physics, New York University, New York, NY 10003, USA\\
$^{15}$Department of Astronomy, University of California at Berkeley, Berkeley, CA 94720, USA\\
}

\def\gs{\mathrel{\raise1.16pt\hbox{$>$}\kern-7.0pt %
\lower3.06pt\hbox{{$\scriptstyle \sim$}}}}         %
\def\ls{\mathrel{\raise1.16pt\hbox{$<$}\kern-7.0pt %
\lower3.06pt\hbox{{$\scriptstyle \sim$}}}}         %

\newcommand{\hompc}{\,h\,{\rm Mpc}^{-1}}
\newcommand{\mpcoh}{\,h^{-1}\,{\rm Mpc}}

\begin{document}

\maketitle

\begin{abstract}
{We present the distance measurement to $z=0.32$ using the 11th data release of the Sloan Digital Sky Survey-III Baryon Acoustic Oscillation Survey (BOSS). We use 313,780 galaxies of the low-redshift (LOWZ) sample over 7,341 square-degrees to compute $D_V=(1264 \pm 25) (r_d/r_{d,fid})$ - a sub 2\%  measurement - using the baryon acoustic feature measured in the galaxy two-point correlation function and power-spectrum. We compare our results to those obtained in DR10. We study observational systematics in the LOWZ sample and quantify potential effects due to photometric offsets between the northern and southern Galactic caps. We find the sample to be robust to all systematic effects found to impact on the targeting of higher-redshift BOSS galaxies, and that the observed north-south tensions can be explained by either limitations in photometric calibration or by sample variance, and have no impact on our final result. Our measurement, combined with the baryonic acoustic scale at $z=0.57$, is used in \cite{Aardwolf} to constrain cosmological parameters.}
\end{abstract}


\title{Clustering in LOWZ BOSS galaxies}

\section{Introduction}  \label{sec:intro}

Current observational evidence increasingly points towards a scenario where the Universe is undergoing an accelerated expansion (see e.g. \citealt{RiessEtAl98, PerlmutterEtAl99, KesslerEtAl09, AmanullahEtAl10, PercivalEtAl10, ReidEtAl10,BlakeEtAl11BAOb, BlakeEtAl11,BlakeEtAl11BAO,  ConleyEtAl11, Aadvark}). The physical reason behind such an acceleration remains a mystery, and potential explanations range from a simple cosmological constant or vacuum density, to modified gravity models or an inhomogeneous Universe creating the illusion of an acceleration. A key goal of modern cosmology, therefore, is to measure the expansion rate of the Universe with increasing precision, with the clear intent of glimpsing the physics behind the Universe's acceleration by direct comparison of this measurement with predictions arising from different physical models. This increase in precision must be matched by an increase in accuracy, and both aspects are a challenge for modern-day galaxy redshift surveys.

The last 30 years have seen a phenomenal increase in our ability to measure the detailed large-scale structure of the Universe via galaxy redshift surveys, e.g. the CfA Redshift Survey \citep{GellerAndHuchra89}, the 2dF Galaxy Redshift Survey \citep{Colless98}, the 6dF Galaxy Survey \citep{JonesEtAl04}, the Sloan Digital Sky Survey \citep{YorkEtAl00}, the VIMOS Public Extragalactic Redshift Survey \citep{GuzzoEtAl13} or the WiggleZ Dark Energy Survey \citep{DrinkwaterEtAl10}. Such surveys provide a wealth of information on cosmological models and on the evolution of galaxies, and are a remarkably versatile tool of modern Astronomy. With enough volume surveyed, the expansion history of the Universe can be measured via the baryon acoustic oscillation (BAO) scale - an imprint of the comoving sound-horizon size at the time of recombination on the distribution of galaxies. The BAO scale has now been convincingly measured using a variety of datasets and methodologies (e.g. \citealt{ColeEtAl05, EisensteinEtAl05, PercivalEtAl07, Aadvark, Abalone}) and, as it can be found at scales $\approx 100 \mpcoh$, it remains free of many astrophysical systematics, providing one of the most robust probes of the expansion history of the Universe (see \citealt{WeinbergEtAl12} for a review). Nonetheless, as  distance measurements approach a precision of $1\%$, significant pressure is put on our understanding of the data and all aspects of the survey - see \cite{RossEtAl12} for a study on potential large-scale systematic effects in galaxy redshift surveys, and \cite{VargasManagaEtAl14} for a detailed study on fitting the BAO scale to the anisotropic correlation function. 

The Baryon Oscillation Spectroscopic Survey (BOSS, \citealt{DawsonEtAl13}), part of the Sloan Digital Sky Survey-III (SDSS-III, \citealt{EisensteinEtAl11}) is a state-of-the-art experiment surveying an unprecedented volume of the Universe with such high density by targeting 1.5 million galaxies over 10,000 sq. degrees of sky in the redshift range $0.1 < z < 0.8$. Using data-release 9 (DR9, \citealt{AhnEtAl12}), \cite{Aadvark} presented the first BAO measurements from BOSS, achieving a 1.7 per-cent measurement on $D_V (z=0.57) \equiv [cz(1+z)^2D_A^2H^{-1}]^{1/3}$, where $D_A$ is the angular diameter distance and $H$ the Hubble parameter. \cite{Aardwolf} using DR11 (internal data-release), which  corresponds to an increase in the effective area over DR9 of a factor of roughly 2.5, achieve a 1 per-cent measurement of $D_V$ over the same redshift range.

In this paper, we focus on the low-redshift (LOWZ) sample of BOSS and present for the first time robust measurements of the large-scale two-point correlation function $\xi(s)$ and spherically-averaged power-spectrum $P(k)$ at $\langle z \rangle=0.32$ using DR10 (Ahn et al. 2013) and DR11 BOSS data. The LOWZ part of the survey has lagged behind the rest of the survey due to an initial error in target selection, and it did not constitute a competitive sample at $z\approx 0.3$ at the time of DR9. This is no longer the case and in this paper we present the sample, an analysis of potential systematics, clustering measurements and the cosmic scale distances to $z=0.32$ that are used in \cite{Aardwolf} to subsequently constrain cosmological models. 

This paper is organised as follows. In Section~\ref{sec:data} we present our dataset, including target-selection criteria and catalogue creation; in Section~\ref{sec:systematics} we analyse the sample for any  potential systematics; in Section~\ref{sec:analysis} we detail the methods used to make the clustering statistics, including how the reconstruction technique is used to improve our errors; in Section~\ref{sec:BAO_fit} we present our method to fit the BAO scale, and we show robustness tests of our methodology performed on the mocks; in Section~\ref{sec:results} we present and discuss our results and we conclude in Section~\ref{sec:conclusions}. We use a fiducial flat cosmology of $\Lambda$CDM, with $\Omega_m = 0.274$, $h =0.7$, $\Omega_b h^2= 0.0224$ and $\sigma_8=0.8$, matching those used in \cite{Aardwolf}.

\section{Data}\label{sec:data}

BOSS is a spectroscopic survey, targeted from SDSS-III data-release 8 imaging (DR8, \citealt{AiharaEtAl11}) that was designed to map out the large-scale structure of the Universe over an unprecedented volume. All imaging was collected on the 2.5 Sloan Foundation Telescope \citep{GunnEtAl06}, at Apache Point Observatory, New Mexico. A drift-scan mosaic CCD camera \citep{GunnEtAl98} was used to image the sky in five photometric bands: $u$, $g$, $r$, $i$ and $z$ \citep{FukugitaEtAl96, SmithEtAl02, DoiEtAl10}, to a limiting magnitude of $r\approx 22.5$. All magnitudes in this paper have been corrected for Galactic extinction using the dust maps of \cite{SchlegelEtAl98}. For further details spectroscopic and photometric data-processing we refer the reader to \cite{PierEtAl03} (astrometric calibration), \cite{PadmanabhanEtAl98} (photometric calibration) and \cite{BoltonEtAl12} (spectral classification and redshift measurements). We analyse data release 10 (DR10, \citealt{AhnEtAl13}) and data release 11; the latter is a collaboration-only release to be made public with the final data release. Our final results are based on DR11, though we will show DR10 results throughout for completeness, systematic checks and reproducibility. DR11 covers 7,341 square degrees of sky over two Galactic caps. These caps are completely disjoint in the sky, leading to some issues with photometric calibration and are therefore treated separately in our analysis (see Section~\ref{sec:offsets} for full details).

Using a 1000-object fibre-fed spectrograph \citep{SmeeEtAl13}, BOSS is targeting objects to a number density of $3\times10^{-4}h^3$Mpc$^{-3}$. Upon completion, BOSS will have obtained precise three-dimensional positions for $~1.35$ million galaxies over 10,000 deg$^2$ and spanning a redshift range of $0.1 < z < 0.8$. 

\subsection{Target selection}

\begin{figure}
\begin{center}
\includegraphics[width=3in]{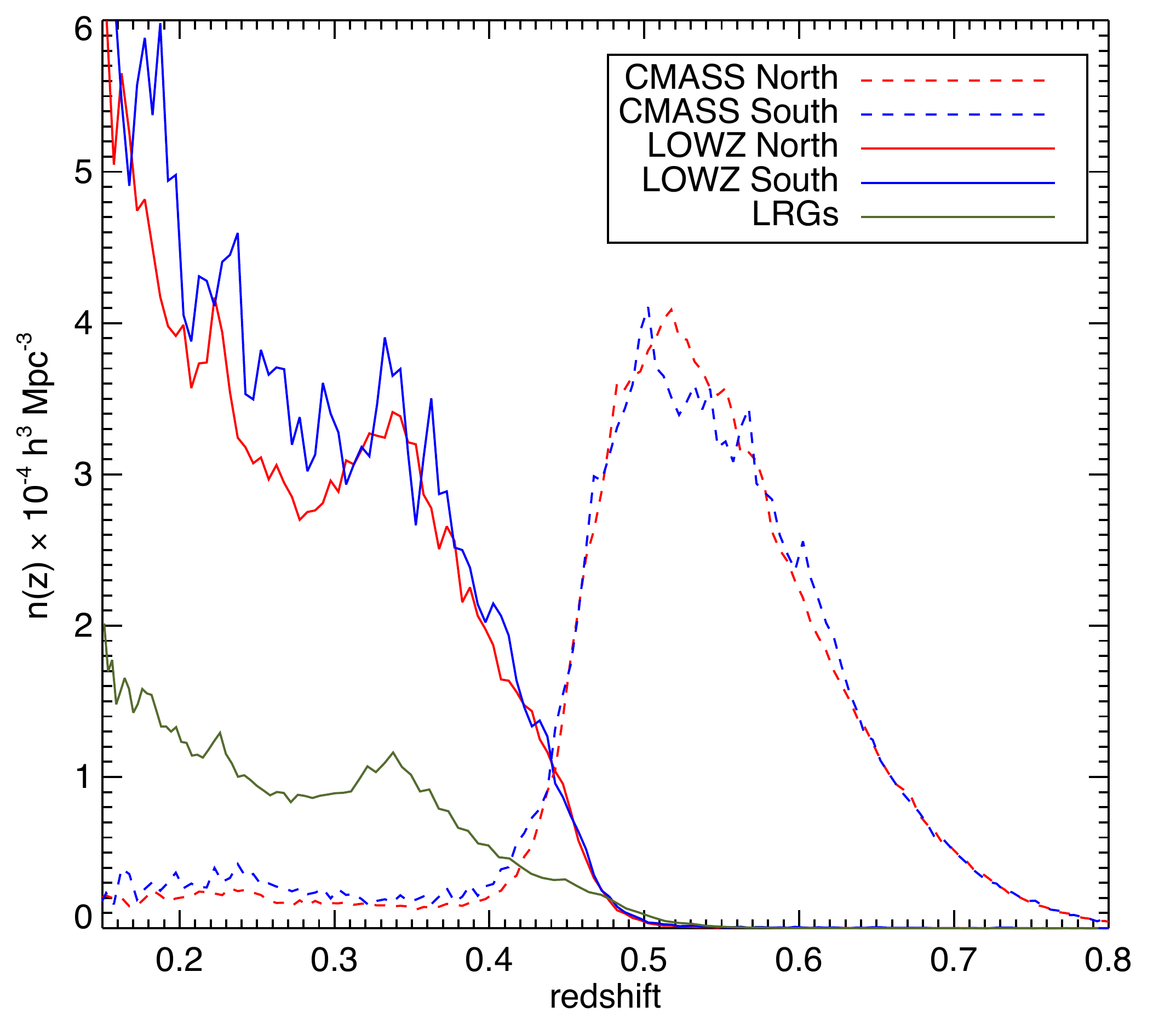}
\caption{The expected number density of BOSS and SDSS-II LRG targets as a function of redshift. The number densities shown here are corrected for missing targets due to close pairs, redshift failures and completeness 
 - see Table~\ref{tab:sample_summary} for details.} 
\label{fig:nz_allsamples}
\end{center}
\end{figure}

BOSS targets objects via two distinct samples: the CMASS (for Constant MASS) sample at $0.4 \lesssim z \lesssim 0.8$, and the LOWZ (for low-redshift) sample at $0.1 \lesssim z \lesssim 0.45$ - see Fig.~\ref{fig:nz_allsamples} for the redshift distributions of these samples. This paper will focus exclusively on the LOWZ sample - for clustering results on the CMASS sample see the companion paper \cite{Aardwolf}.

The LOWZ target selection follows closely the target selection algorithm designed for Luminous Red Galaxies (LRGs) in SDSS-I/II, and described in \cite{EisensteinEtAl01}. The LOWZ sample has approximately three times the number density of the LRG sample (see Fig~\ref{fig:nz_allsamples}), achieved by targeting fainter targets whilst keeping similar colour cuts. Explicitly, a LOWZ galaxy must pass the following conditions:

\begin{eqnarray}
  r_{cmod} &<& 13.6 + c_\parallel/0.3, \label{eq:sliding_cut}\\
  |c_\perp| &<& 0.2, \label{eq:track_cut}\\
  16 < r_{cmod} &<& 19.6,\\
  r_{psf} - r_{cmod} &>& 0.3,
\end{eqnarray}
where the two auxiliary  colours $c_\parallel$ and $c_\perp$ are defined as:
\begin{eqnarray}
c_\parallel &=& 0.7(g-r) + 1.2[(r-i) - 0.18] \\
 c_\perp &=& (r-i) - (g-r)/4 - 0.18.
\end{eqnarray} 

All colours are computed using SDSS model magnitudes, and we use the subscript $psf$ and $cmod$ to denote point-spread function (PSF) and cmodel magnitudes, respectively. 

This is the current (and final) target selection algorithm for LOWZ galaxies. However, due to an error, a different target selection was implemented for the first nine months of BOSS observations. As a result, in order to select a uniform sample of LOWZ galaxies one must only include the regions with TILE$\ge 10324$ (see also \citealt{ParejkoEtAl13} and the detailed target-selection notes in \url{http://www.sdss3.org/dr9/algorithms/boss_galaxy_ts.php}). For this reason, the total area of the LOWZ footprint lags behind that of CMASS. In DR11, the total area of the LOWZ sample is 7,562 sq degrees (see Table~\ref{tab:sample_summary}), whereas for CMASS it is 8,498 sq degrees (see \citealt{Aardwolf})

A first analysis of the LOWZ galaxy sample was presented in \cite{ParejkoEtAl13}, where an analysis of the small-scale clustering of this sample showed that LOWZ galaxies are highly biased (galaxy bias $b\sim 2.0$), that they occupy massive haloes (with an average halo mass of $5.2 \times 10^{13}$h$^{-1}$M$_\odot$), and that they lie at the top end of the stellar mass function with a typical stellar mass of $10^{11.2}$M$_\odot$. Their satellite fraction is measured to be low ($\approx 12.2 \pm 2$ per cent), and they have redder rest-frame colours, on average, than the CMASS sample. To a good approximation one can think of the LRG sample as being a subset of the LOWZ sample, which mostly extends to lower luminosities, stellar and halo masses and galaxy bias. 
\begin{figure*}
\begin{center}
\includegraphics[width=3.3in]{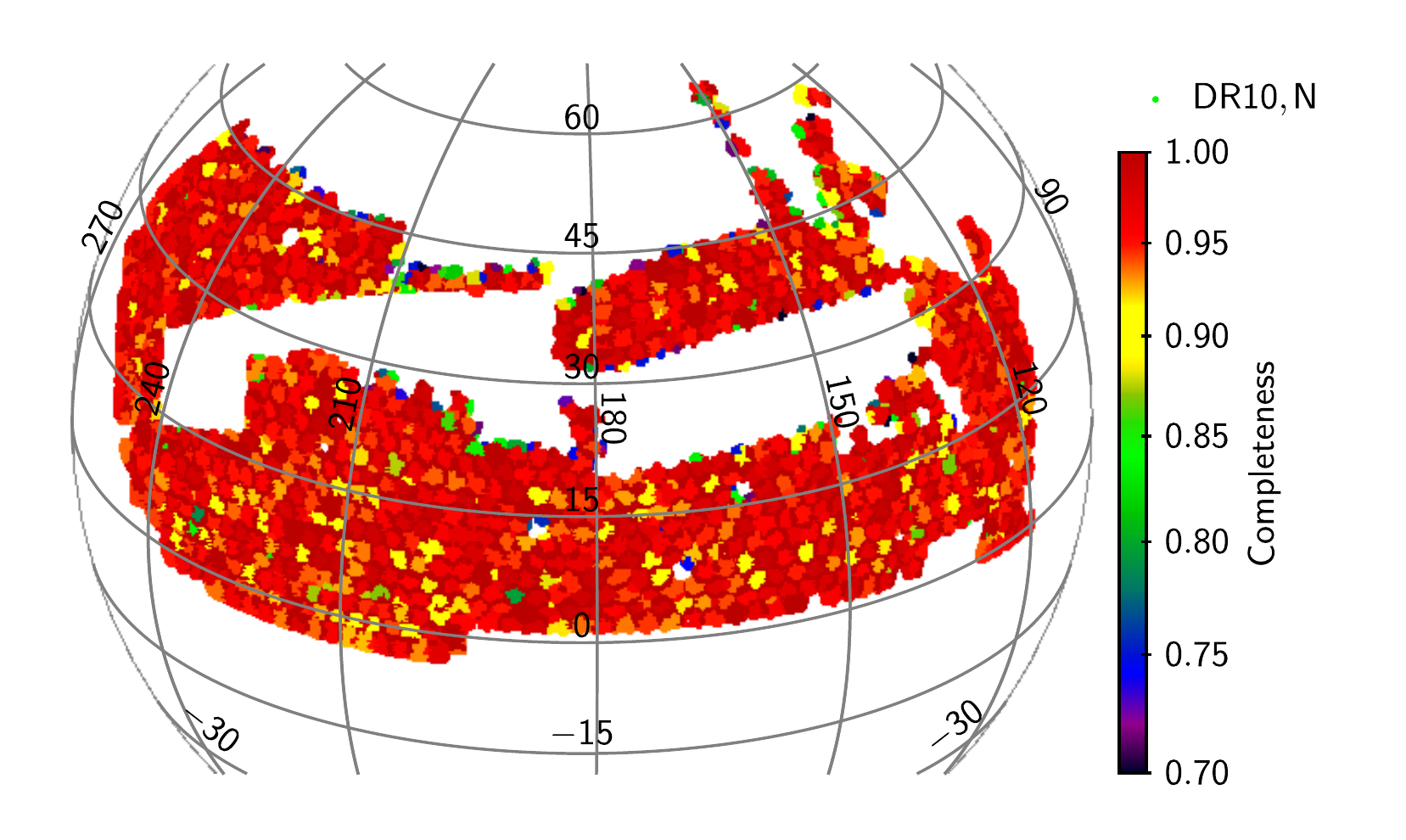}
\includegraphics[width=3.3in]{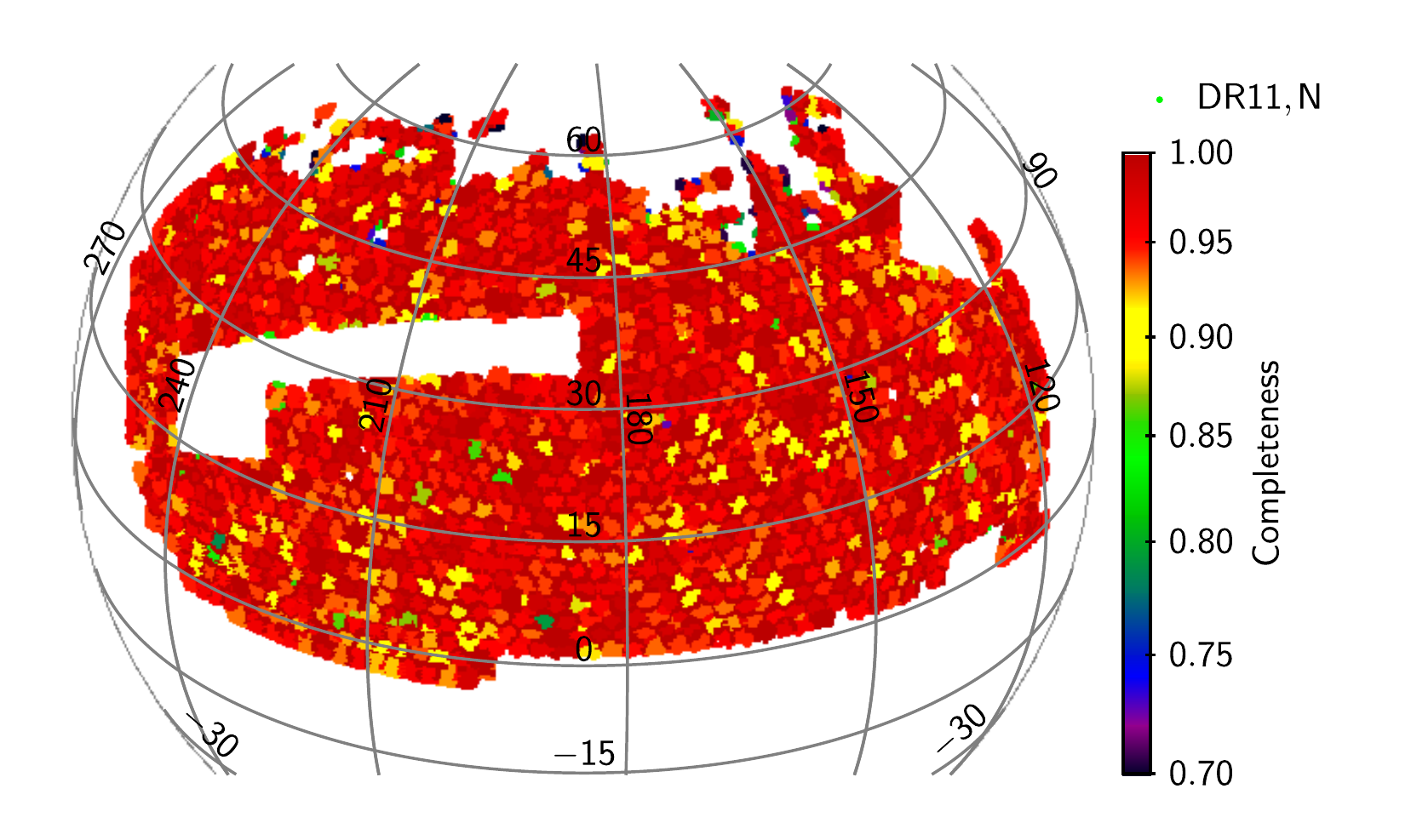}
\includegraphics[width=3.3in]{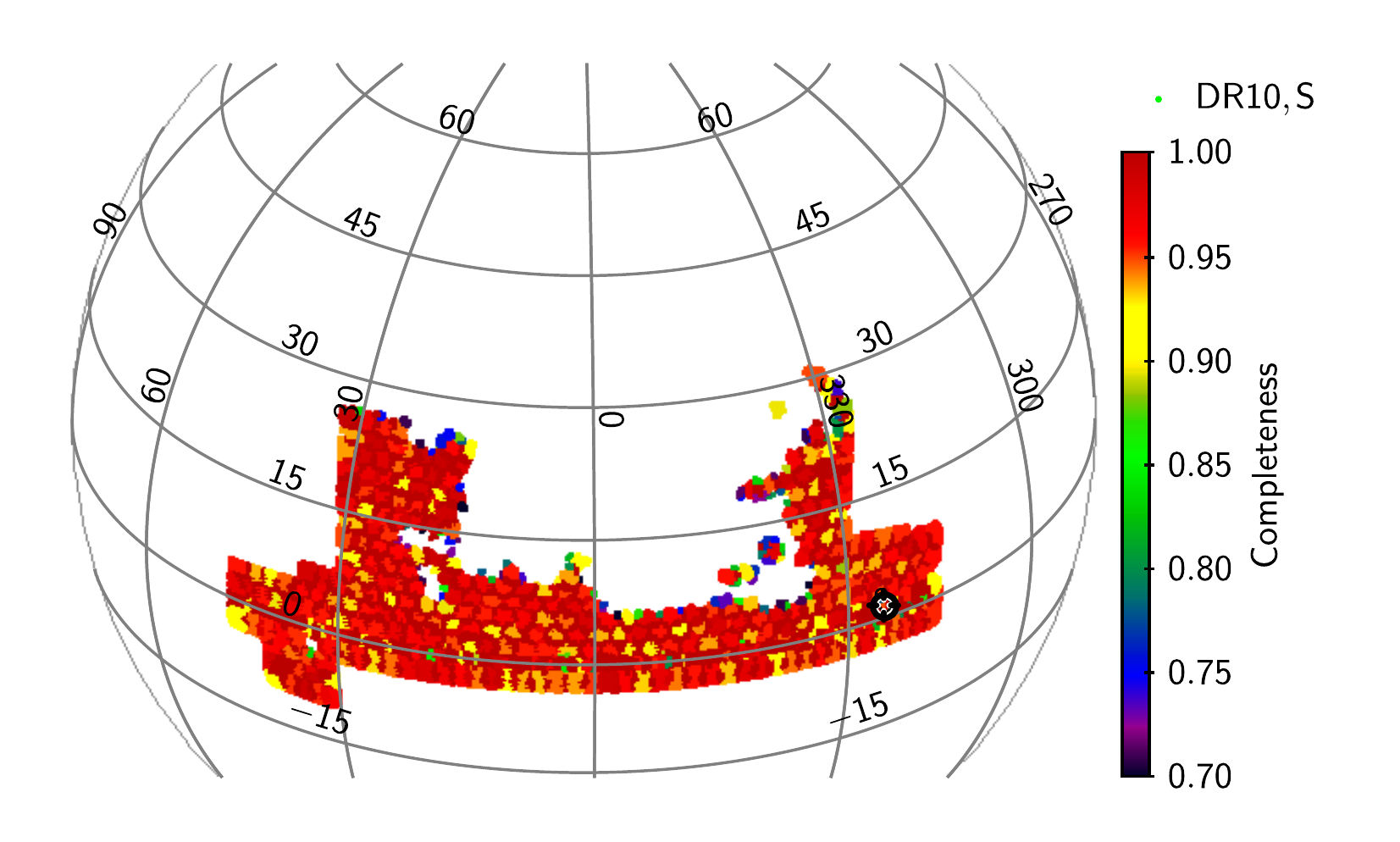}
\includegraphics[width=3.3in]{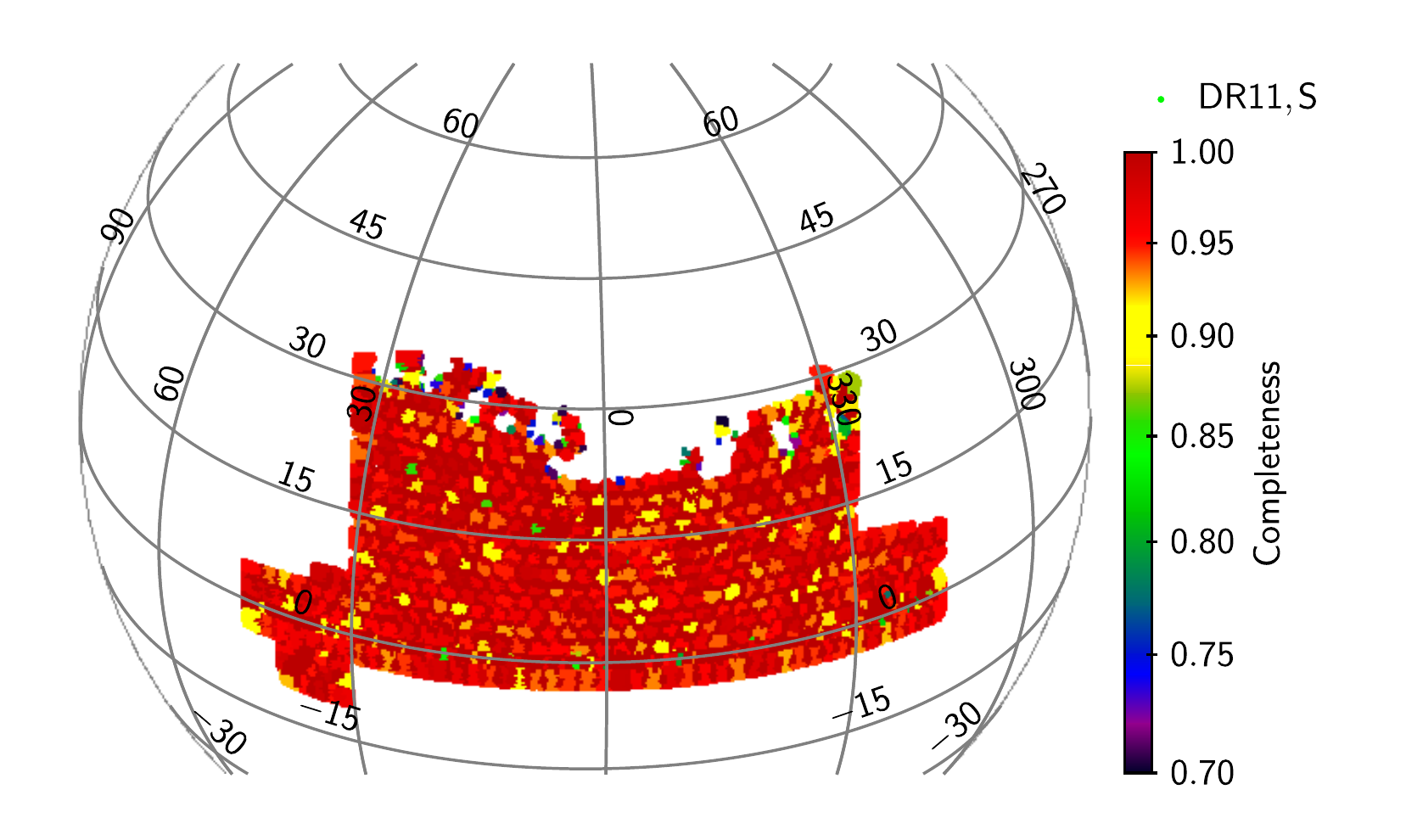}
\caption{Survey footprint in equatorial coordinates. DR10 on the two left panels and DR11 on the two right panels (North and South galactic caps in the top and bottom panels, respectively). The color code shows the completeness of each sector.} 
\label{fig:DR10_footprint}
\end{center}
\end{figure*}

\subsection{Galaxy catalogues}\label{sec:catalogue}

A catalogue suitable for the analysis of large-scale structure was constructed following the procedures detailed in \cite{Aardwolf}. The only exception concerns the treatment of systematic errors, which we address in the next section.

Completeness is estimated as in \cite{Aardwolf}. Corrections for close-pairs (targets that could not be observed due to fibre collisions) and redshift failures (targets for which a spectrum was taken, but no redshift could be measured) are addressed by up-weighting the nearest galaxy with a successful redshift measurement by one unit for each missed target (see  \citealt{Aardwolf}). We denote these weights by $w_{cp}$ and $w_{rf}$ respectively. We also implement the weighting scheme of \cite{FeldmanEtAl94}, to optimally balance the effects of shot-noise and sample variance ($w_{FKP} = 1/[ 1+\bar{n}(z)P_0] $, with $P_0 = 20000h^{-3}$Mpc$^3$). Objects in the catalogue are therefore weighted by a combined weight given by

\begin{equation}
w_{tot} = w_{FKP}(w_{rf} + w_{cp} + 1).
\end{equation}

Table~\ref{tab:sample_summary} summarises the LOWZ DR10 and DR11 samples, and Fig.~\ref{fig:DR10_footprint} shows the survey footprints in the southern and northern Galactic caps (SGC and NGC, respectively).

\begin{table*}
\caption{Basic properties of the LOWZ target class and corresponding mask as defined in the text.}
\label{tab:sample_summary}
\begin{center}
\begin{tabular}{lrrrrrr}
\hline
\hline
 & \multicolumn{3}{c}{DR10} & \multicolumn{3}{c}{DR11}\\
 Property & NGC & SGC & total  & NGC & SGC & total\\ \hline
 $\bar{N}_{\rm gal}$ &113,624 & 67,844 & 181,468 & 156,569 & 108,800 & 265,369 \\
$\bar{N}_{\rm known}$ &89,989 & 8,959 & 98,948 & 124,533 & 11,639 & 136,172 \\
$\bar{N}_{\rm star}$ &804 & 523 & 1,327 & 944 & 754 & 1,698 \\
$\bar{N}_{\rm fail}$ &477 & 278 & 755 & 726 & 497 & 1,223 \\
$\bar{N}_{\rm cp}$ &8,199 & 2,928 & 11,127 & 10,818 & 4,162 & 14,980 \\
$\bar{N}_{\rm missed}$ &7,148 & 2,420 & 9,568 & 9,089 & 3,272 & 12,361 \\
$\bar{N}_{\rm used}$ &157,869 & 61,036 & 218,905 & 219,336 & 94,444 & 313,780 \\
$\bar{N}_{\rm obs}$ &114,905 & 68,645 & 183,550 & 158,239 & 110,051 & 268,290 \\
$\bar{N}_{\rm targ}$ &220,241 & 82,952 & 303,193 & 302,679 & 129,124 & 431,803 \\
Total area / deg$^2$ &4,205 & 1,430 & 5,635 & 5,793 & 2,205 & 7,998 \\
Veto area / deg$^2$ &252 & 58 & 309 & 337 & 99 & 436 \\
Used area / deg$^2$ &3,954 & 1,372 & 5,326 & 5,456 & 2,106 & 7,562 \\
Effective area / deg$^2$ &3,824 & 1,332 & 5,156 & 5,291 & 2,051 & 7,341 \\
\hline
\end{tabular}
\end{center}
\end{table*}


In this paper we use galaxies with $0.15 < z < 0.43$. At lower redshifts the target selection of LOWZ galaxies is increasingly contaminated by objects that differ from the typical LRG, explaining the rapid increase in number density (see also \citealt{EisensteinEtAl01, TojeiroEtAl11}). The high redshift limit was introduced to ensure no overlap between the LOWZ and CMASS samples, and was chosen where the galaxy densities were approximately equal. The final DR11 LOWZ sample contains 313,780 galaxies, over an effective area of 7,341 sq. degrees.

\subsection{Random catalogues}\label{sec:randoms}

We construct random catalogues that mimic the angular and radial selection function of the SGC and NGC independently, with a density that is 30 times that of the data. We decouple the radial and angular components and generate random angular positions within the angular mask of the survey whilst assigning a random galaxy redshift to each random point (this corresponds to the "shuffled" technique described in \citealt{SamushiaEtAl12, RossEtAl12}). Finally we subsample the random catalogue to  match the on-sky completeness of the data (see Section~\ref{sec:catalogue}).

\section{Checking for systematic effects}\label{sec:systematics}

In this section we present our investigation into potential systematic effects on the LOWZ sample. We begin by investigating the effects of stellar density, airmass, extinction and seeing on LOWZ targets. Then examine the potential effects of photometric offsets between the two Galactic caps on number density and type of target. Finally we investigate a difference between the observed clustering in the two Galactic caps and its significance. We show that the LOWZ sample is immune to many of the issues found with the CMASS sample, and that most of the discrepancies found are consistent with what we expect from sample variance as estimated from mocks (see Section \ref{sec:mocks}). We do not apply any systematic weights to the LOWZ galaxies. 

\subsection{Observational effects on targets}

\cite{RossEtAl12} showed how observational effects such as stellar density, airmass, seeing or sky brightness affect the large-scale power of CMASS targets. The main effect was demonstrated to be due to obscuration of CMASS targets due to stars, a signature which imparted a large-scale signal in the two-point correlation function of CMASS galaxies. A corrective weighting scheme was developed, which ensured that weighted angular target-density showed no significant dependence on stellar density. 

\begin{figure}
\begin{center}
\includegraphics[width=3in]{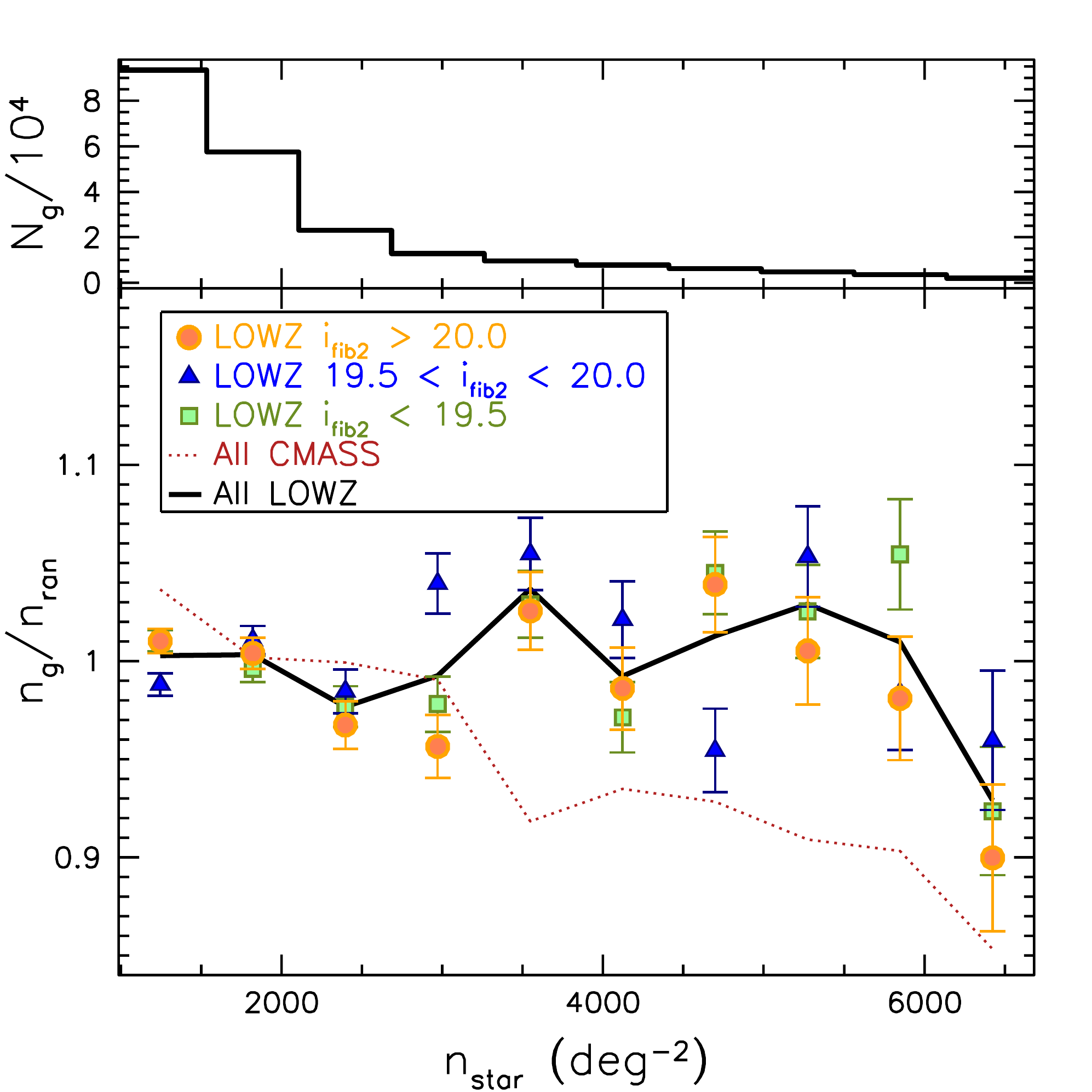}
\caption{{\it Bottom panel:} Fluctuations of on-sky angular target density in the NGC as a function of angular stellar density. The coloured symbols show this relationship for LOWZ targets within three $i_{fib2}$ bins, and the thick black line represents the relationship as measured on the whole LOWZ sample. Error bars show Poisson errors. Unlike what is observed for the fainter CMASS sample (see \protect\citealt{RossEtAl12}), we see no coherent trend of target density with stellar density. {\it Top panel:} Distribution of LOWZ targets according to angular stellar density.} 
\label{fig:LOWZvst}
\end{center}
\end{figure}

This effect was shown to depend on fibre-magnitude and to be significant only for CMASS galaxies with $i_{fib2} \gtrsim 20.5$. Fewer than 2 per-cent of the LOWZ targets have $i_{fib2} > 20.5$, and we thereby expect the LOWZ sample to be immune to such effects. Fig.~\ref{fig:LOWZvst} displays angular target fluctuations of the LOWZ sample with stellar density for galaxies of different  $i_{fib2}$ brightness - as expected we find no evidence that stellar density affects the target density of LOWZ targets (see also Fig. 11 of \citealt{RossEtAl12}), even at the faint end.

\begin{figure*}
\centering
  \resizebox{0.95\textwidth}{!}{\includegraphics{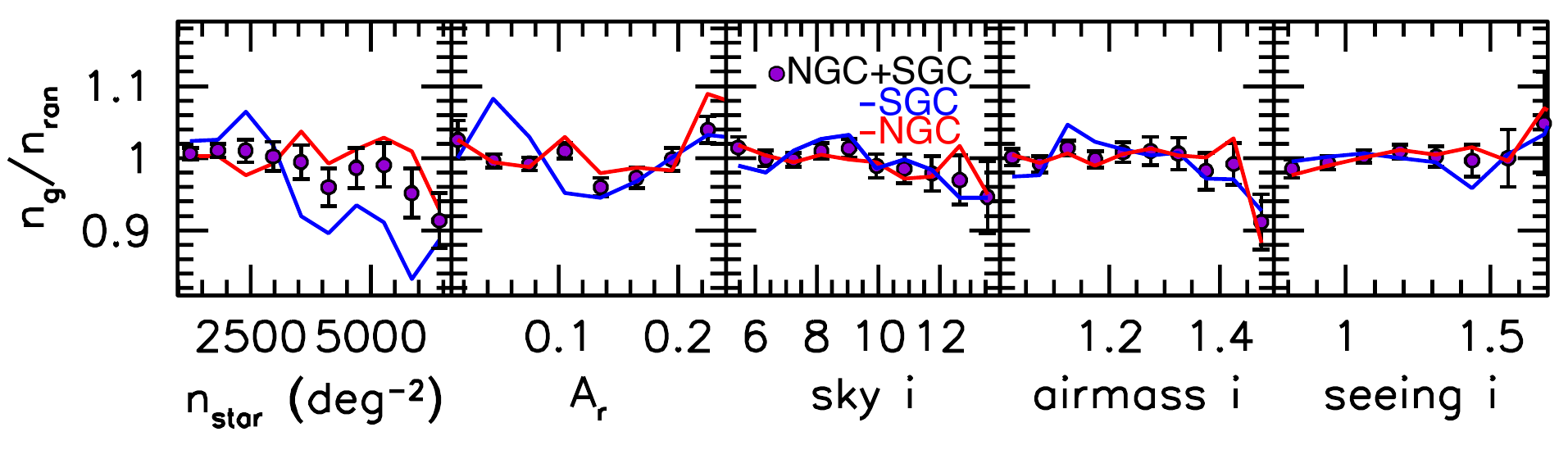}}
    \caption{Fluctuations of on-sky angular target density for the NGC (red lines), SGC (blue lines) and their combination (purple circles), as a function of angular stellar density, $r$-band extinction, $i$-band sky background, airmass in the $i$ band, and seeing in the $i$ band, from left to right. No significant trends are apparent.} 
\label{fig:LOWZvstNS}
\end{figure*}

We find a more significant trend of target density with stellar density in the SGC than in the NGC, as shown in the first panel of Fig.~\ref{fig:LOWZvstNS}. Coincidentally, there is a difference in the large-scale amplitude of the correlation function between the two Galactic caps (see Fig.~\ref{fig:xi0_NS} and Section ~\ref{sec:clustering_sys}). However, no weighting based on stellar density, airmass, seeing or sky brightness significantly alleviates this tension. A detailed investigation of the differences between the NGC and SGC large-scale power is presented in Section~\ref{sec:clustering_sys}, where we conclude that the observed differences are not unusual given the expectation from the mocks, and are indeed alleviated in DR11 when compared to DR10.

We thereby apply no systematic weights based on stellar density, airmass, seeing or sky brightness to the LOWZ sample.

\subsection{Photometric offsets}\label{sec:offsets}

A potential offset in photometric calibration between the NGC and the SGC was first reported in \cite{SchlaflyEtAl10}. These two regions are completely disjoint in the sky (see Fig.~\ref{fig:DR10_footprint}), and thereby cross-calibration is particularly difficult. Using the blue tip of the stellar locus as a standard colour, the authors found the SGC to have systematically redder colour in $g-r$, $r-i$ and $i-z$ by, on average, 21.8, 7.2 and 12.4 mmag, respectively. However, the blue-tip method is sensitive to stellar population parameters, such as metallicity, thereby making it impossible to disentangle photometric offsets from an intrinsically different stellar population in the South. In \cite{SchlaflyEtAl11} the authors used stellar spectra to predict broadband colours of stars, using detailed fits to their spectra that allows one to solve for stellar parameters such as metallicity, temperature and gravity. Overall they confirm the results found using the blue-tip method, but revised the colour offsets between the NGC and the SGC to be $8.8 \pm 1.5$, $3.4 \pm 1.9$ and $9.3 \pm 1.6$ mmag in $g-r$, $r-i$ and $i-z$, respectively. \cite{RossEtAl11b} used these offsets to explain the observed difference in number density of CMASS galaxies in the two Galactic caps. In this section we investigate the impact on the LOWZ sample.

\subsubsection{Number density}\label{sec:number_density}
All selection cuts in the target algorithm depend on observed magnitudes as well as colours and we have no knowledge of how each individual band is affected  by the photometric offsets. Therefore, we are unable to re-target the SGC and establish whether the difference in $n(z)$ between the two Galactic caps seen in Fig.~\ref{fig:nz_allsamples} can be fully explained by issues with the photometry reported by \cite{SchlaflyEtAl11}. Nonetheless, we expect most of the effect to enter through the $c_\parallel$ cut and Eq.~\ref{eq:sliding_cut} - this is the target selection cut at which the number density of LOWZ targets is highest, and has the highest gradient (see e.g. \citealt{RossEtAl12}). According to Table 5 of \cite{SchlaflyEtAl11}, the offset in $c_{\parallel}$ between the two Galactic caps is 0.01 mag (note: this differs from the value of 0.015 mag quoted in \citealt{RossEtAl12} due to a revision of the colour offsets between the submitted and final version of \citealt{SchlaflyEtAl11}).

\begin{figure}
\begin{center}
\includegraphics[width=3in]{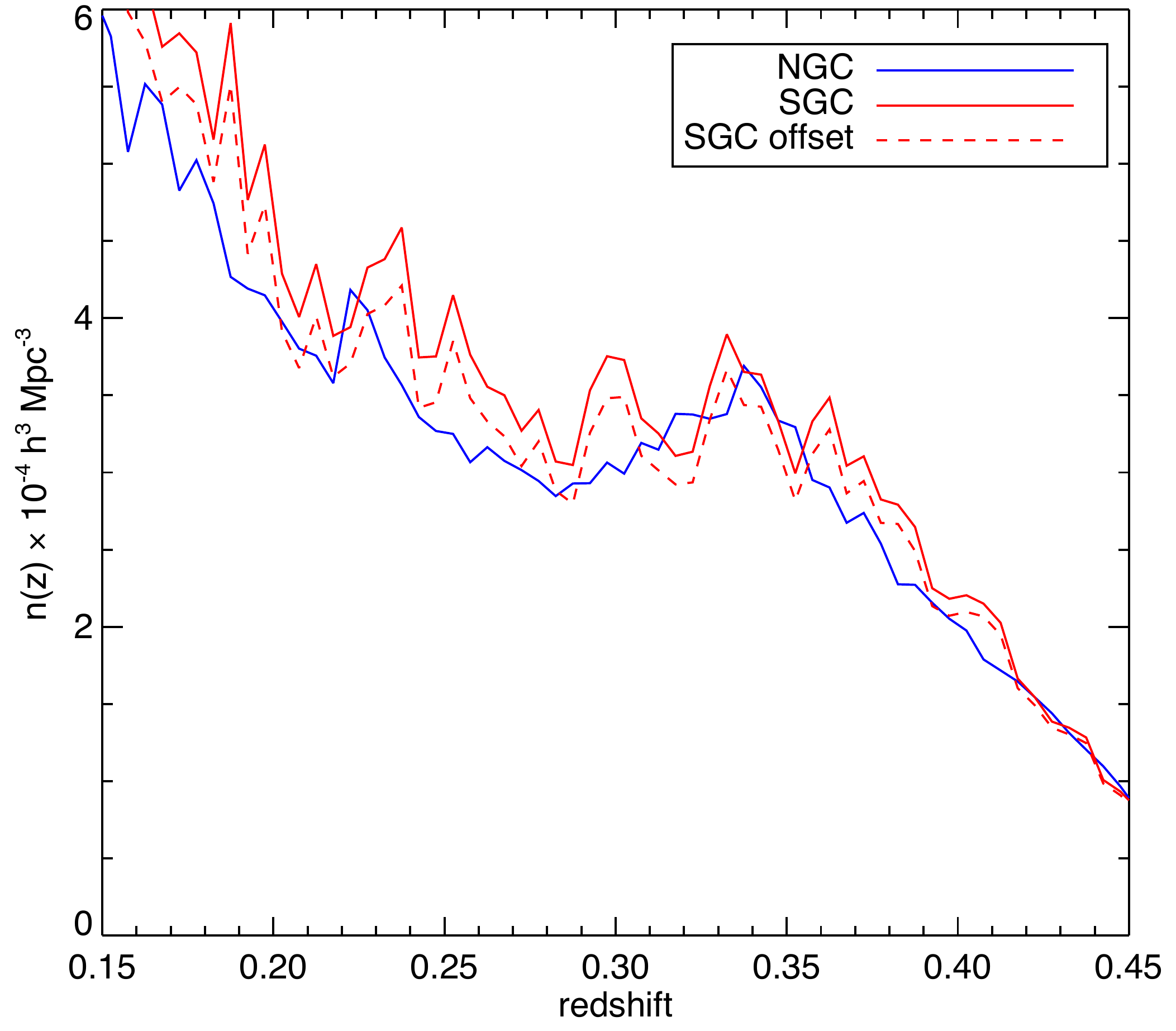}
\caption{Weighted galaxy number density for LOWZ galaxies as a function of redshift for the NGC and SGC (solid lines, lower and upper respectively). The dashed line is the expected number density in the SGC once colour offsets between the two galactic caps, as reported in \protect\cite{SchlaflyEtAl11}, are taken into consideration. The tension between the two galactic caps is significantly alleviated, but it remains unusual at a $3\sigma$ level - see text for details.} 
\label{fig:nbar_schl}
\end{center}
\end{figure}

Applying an offset of 0.01 mag to the $c_\parallel$ colour of all galaxies in the SGC results in a reduction of LOWZ targets that diminishes the tension between NGC and SGC number densities - see Fig.~\ref{fig:nbar_schl}. Using 1000 PTHalo mocks to estimate the covariances in $n(z)$ (see Section~\ref{sec:mocks}), we compute a $\chi^2$ between the NGC and the SGC for DR11. The North and the South, without applying the colour offsets, are incompatible at the $0.0005$ per-cent level, with a $\chi^2 = 73 $ for 28 degrees of freedom - i.e. we expect a $\chi^2$ value larger than what we measure only $0.005$ per-cent of the time. Once we apply the 0.01 mag offset to the $c_\parallel$ colour of the SGC galaxies,  this $\chi^2$ value is reduced to $49$ for 28 degrees of freedom, with a probability of $0.8$ per-cent. Whereas it is clear that the photometric offsets go some way to alleviate the tension between the northern and southern Galactic caps, the level of disagreement remains above the $3\sigma$ level. Most of the contribution to these $\chi^2$ values is produced by the lower redshift bins, which contribute modestly to the overall volume. The discrepancy between the NGC and SGC falls to $2\sigma$ when we only consider the redshift range of $0.25 < z < 0.43$ (representing over 80 per-cent of the survey volume).


The absolute difference in number density between the two hemispheres is similar for DR10 and DR11, and therefore the significance of the offset is larger for DR11 as sample variance is largely reduced in this dataset. This suggests that the observed offset in number density is indeed due to systematics offsets in the data. At the moment we do not have enough information on photometric offsets to determine whether they could explain the {\it full} difference in number density.

Therefore, given the large uncertainty in the computation of colour offsets and their exact effect in the targeted samples, we make no attempt to re-sample the SGC targets to a sample that would be more compatible with the NGC. We instead treat the NGC and SGC as two independent samples, each with its own selection function. We show in the next section that, in spite of this offset, the intrinsic colours of the observed galaxies are quite similar in the two Galactic caps - though they are on average fainter in the SGC as expected. We optimally combine clustering results for the NGC and SGC as detailed in Section~\ref{sec:analysis} to obtain our final measurements. 

\subsubsection{Colours}\label{sec:colours}

\begin{figure}
\includegraphics[width=3in]{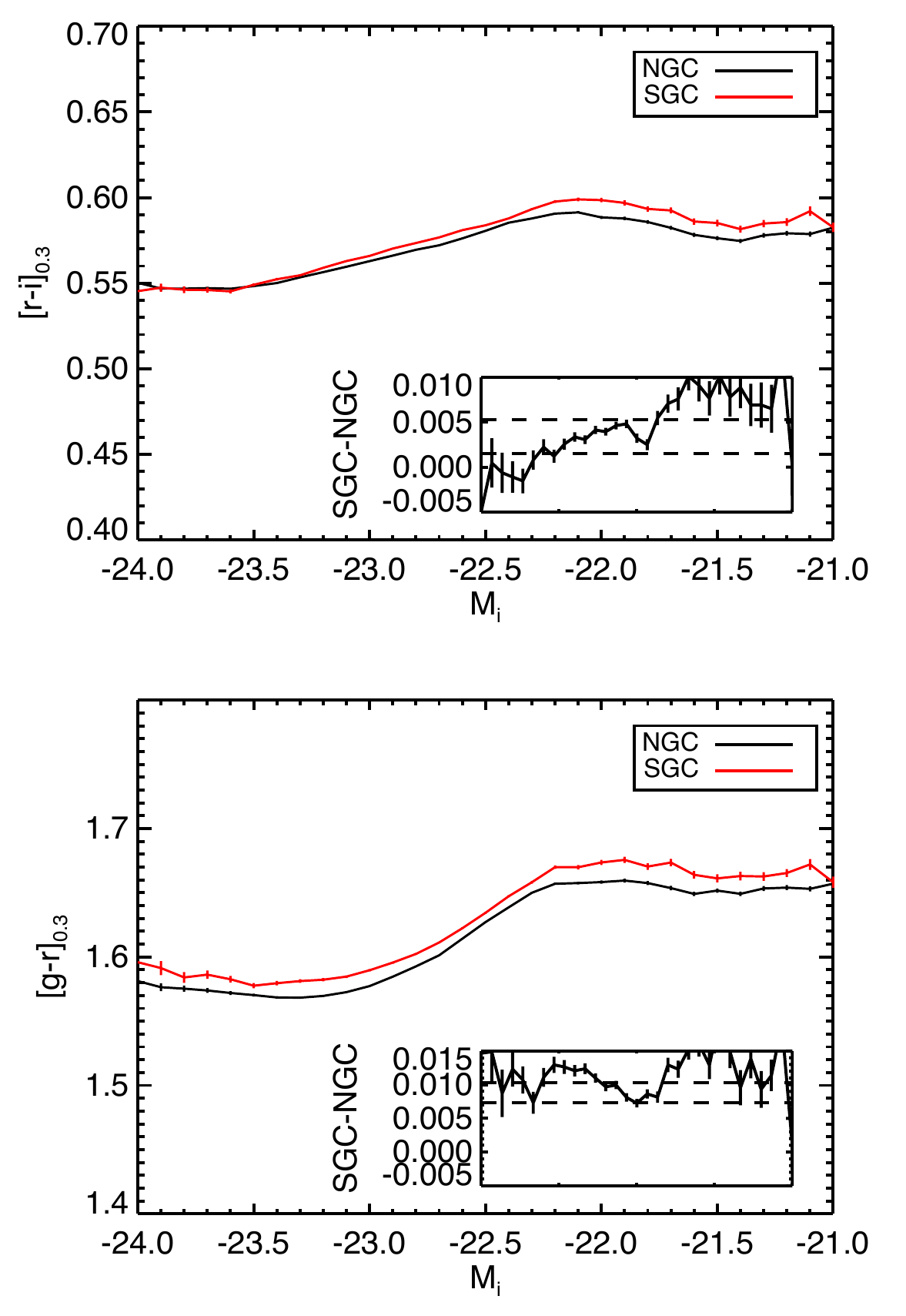}
\caption{ Median rest-frame $r-i$ and $g-r$ colours (k-corrected to $z=0.3$, and computed for filters shifted to the same redshift) as a function of $r-$band absolute magnitude for DR11 NGC (in black) and the DR11 SGC (in red). The inset in each panel shows the difference between the median colour in the two galactic caps, in the same $x-$axis range as in the main panel. The dashed horizontal lines in the insets show the offset in each colour as computed by \protect\cite{SchlaflyEtAl11} (the two lines show the 1-$\sigma$ range). The measured offsets of the median rest-frame colours of the LOWZ galaxies are broadly consistent with the expected offsets of \protect\cite{SchlaflyEtAl11}.}
\label{fig:NS_colours}

\end{figure}

A photometric offset impacts on the sample of observed galaxies in the NGC and SGC as it shifts the underlying distribution of observed colours across the targeting boundaries. We would like to know how this offset impacts on the {\it type} of galaxies that are targeted in each galactic cap.

Comparing the distribution of observed colours and magnitudes of the two samples brings little insight, due to the spread introduced by the $n(z)$ of the samples. Instead we examine intrinsic (rest-frame) colours and magnitudes and their difference between the NGC and SGC. To compute rest-frame colours we use the k-corrections of the purely passive model of \cite{MarastonEtAl09}. Rest-frame colours are computed in shifted filters to $z=0.3$, thereby minimising the dependence on the modelling whilst providing rest-frame colours that can be directly compared to the estimated photometric offsets of \cite{SchlaflyEtAl11}.

 
We present the offset in rest-frame colours between the NGC and SGC in Fig.~\ref{fig:NS_colours}. 
The insets in each panel display the offset in the median colours between the two hemispheres, and how these compare to the offsets of \cite{SchlaflyEtAl11} ($\pm 1\sigma$ values are shown in the dashed lines). 

The offsets in rest-frame $[r-i]_{0.3}$ and $[g-r]_{0.3}$ are in good agreement with the predicted offsets reported by \cite{SchlaflyEtAl11}. This result suggests that the targeting algorithm is selecting similar types of galaxies in the two hemispheres, in spite of the offset in photometry. As the LOWZ targeting was designed to select a well-defined and isolated red-sequence, this is not a surprise. Finally, the SGC galaxies have on average a fainter absolute $r-$band magnitude by $0.01$ mags - a value consistent with the increase in number density in the south being driven by the inclusion of slightly fainter targets, but of similar intrinsic colours.

%

\subsection{Clustering}\label{sec:clustering_sys}

\begin{figure}
\includegraphics[width=3.2in]{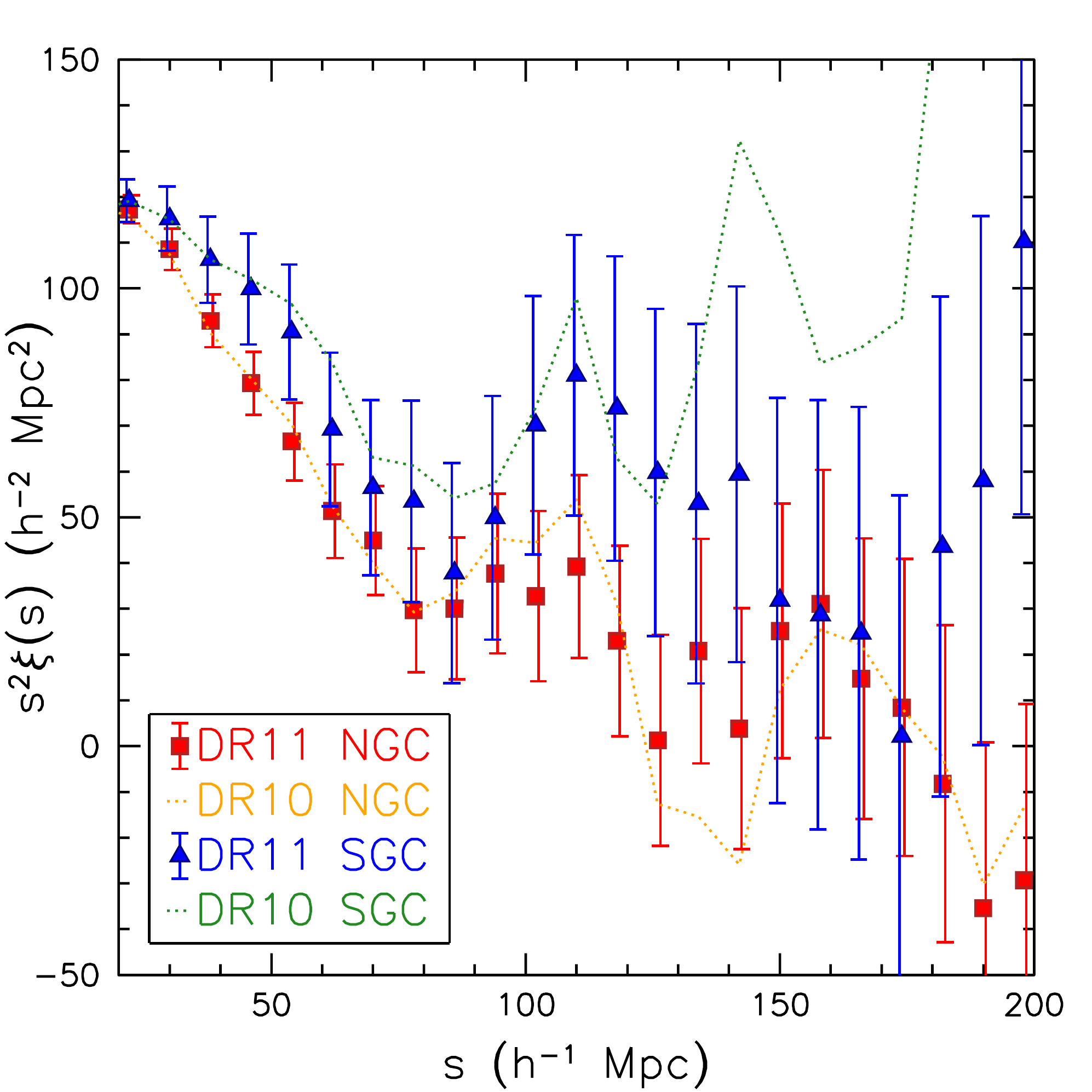}
\caption{The two-point correlation function for the NGC and SGC, pre-reconstruction. The data points show $\xi(s)$ for DR11 data, with error bars computed using the 1000 PTHalo mocks described in Section~\ref{sec:mocks}. The measurements in the two regions are consistent: $\chi^2 = 27.7$ when comparing the 23 data bins in the range $20 < s < 200h^{-1}$Mpc. The dotted lines display the measurements for the DR10 sample, where greater disagreement between the two regions is observed. See the text for more details.}
\label{fig:xi0_NS}
\end{figure}

Fig.~\ref{fig:xi0_NS} shows the two-point correlation function of LOWZ galaxies in the northern and southern Galactic caps, with points displaying the DR11 data and dotted lines presenting the DR10 results. The excess of power observed in the SGC, although clearly visible by eye, is consistent with sample variance. We compute the $\chi_{NS}^2$ between the two hemispheres as

\begin{equation}
\chi^2_{NS} = \sum_{ij} \lbrack \xi_N(s_i) - \xi_S(s_i) \rbrack C_{ij}^{-1} \lbrack \xi_N(s_j) - \xi_S(s_j) \rbrack,
\end{equation}
where the subscripts $_N$ and $_S$ to refer to the NGC and SGC respectively, and $C = C_N + C_S$ as we expect the measurements in the two Galactic caps to be uncorrelated. 

For DR11, we find $\chi^2_{NS} = 27.7$ for 23 dof in the range $20 < s < 200 h^{-1}$ Mpc. A larger $\chi^2$ would be expected 23 per cent of the time, and thereby not unusual.

The results for DR10 are somewhat more discrepant, but still do not raise cause for significant concern: we find $\chi^2_{NS} = 30.9$ for the same 23 bins with $20 < s < 200 h^{-1}$ Mpc, a larger value would be expected 12.5 per cent of the time.

\begin{figure}
\begin{center}
\includegraphics[width=3.2in]{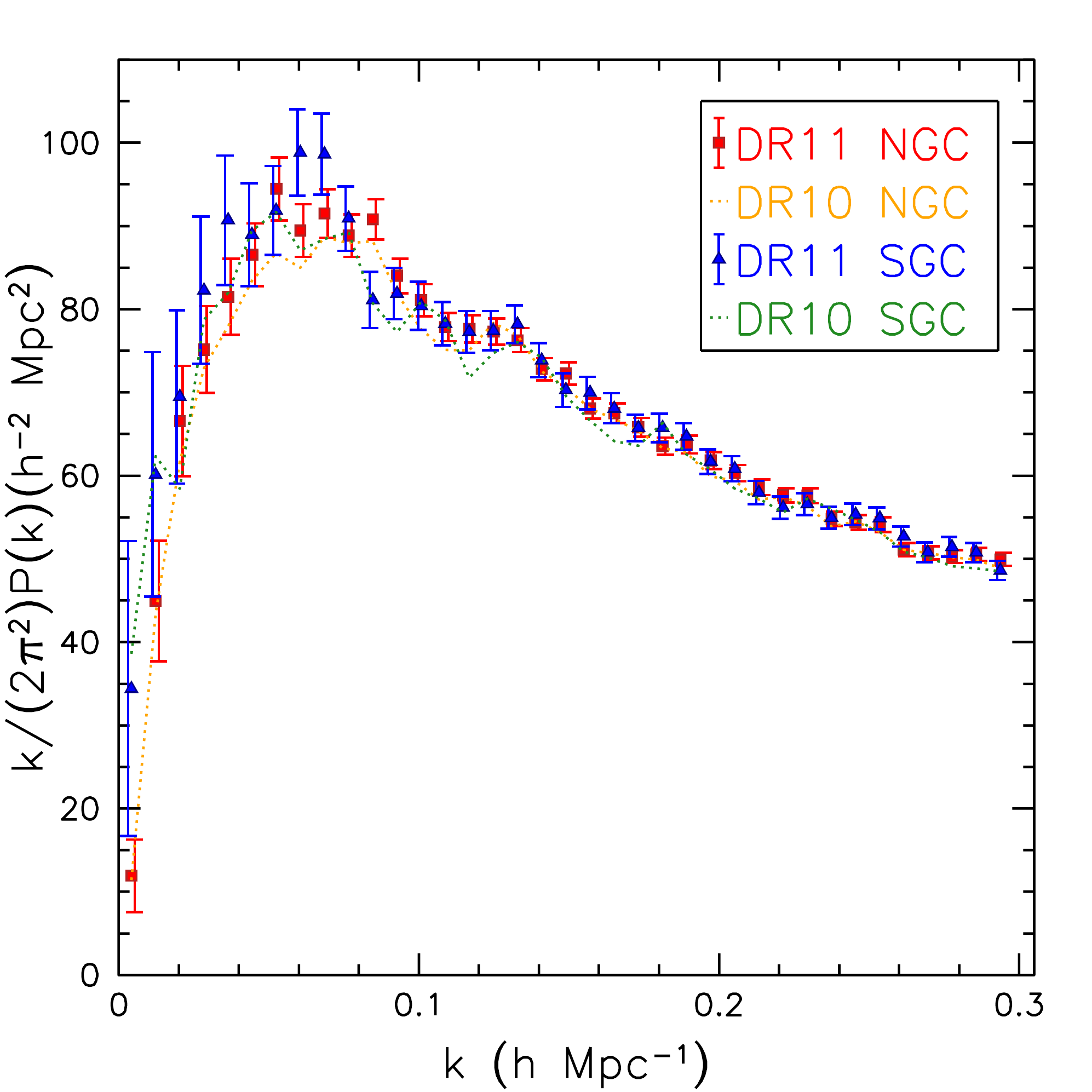}
\caption{ The spherically-averaged power spectrum for the NCG and SGC. The error bars computed using the 1000 DR11 PTHalo mocks described in Section~\ref{sec:mocks}.} 
\label{fig:Pk_NS}
\end{center}
\end{figure}

Fig.~\ref{fig:Pk_NS} shows the spherically averaged power spectra for the NGC and SGC. The two curves appear broadly consistent. The power in the SGC is greater than that of the NGC primarily at $k < 0.06 \mpcoh$. One would expect the power in the SGC to be less than that of the NGC, due to its smaller footprint and therefore larger integral constraint. Due to these window function effects, one cannot directly compare the NGC and SGC $P(k)$ measurements, but the results appear consistent with our findings for $\xi(s)$ that the two regions yield consistent clustering measurements. In Section \ref{sec:baorobust}, we present BAO measurements for each region.

\subsubsection{Fluctuations with Right Ascension}
\begin{figure}
\begin{center}
\includegraphics[width=3in]{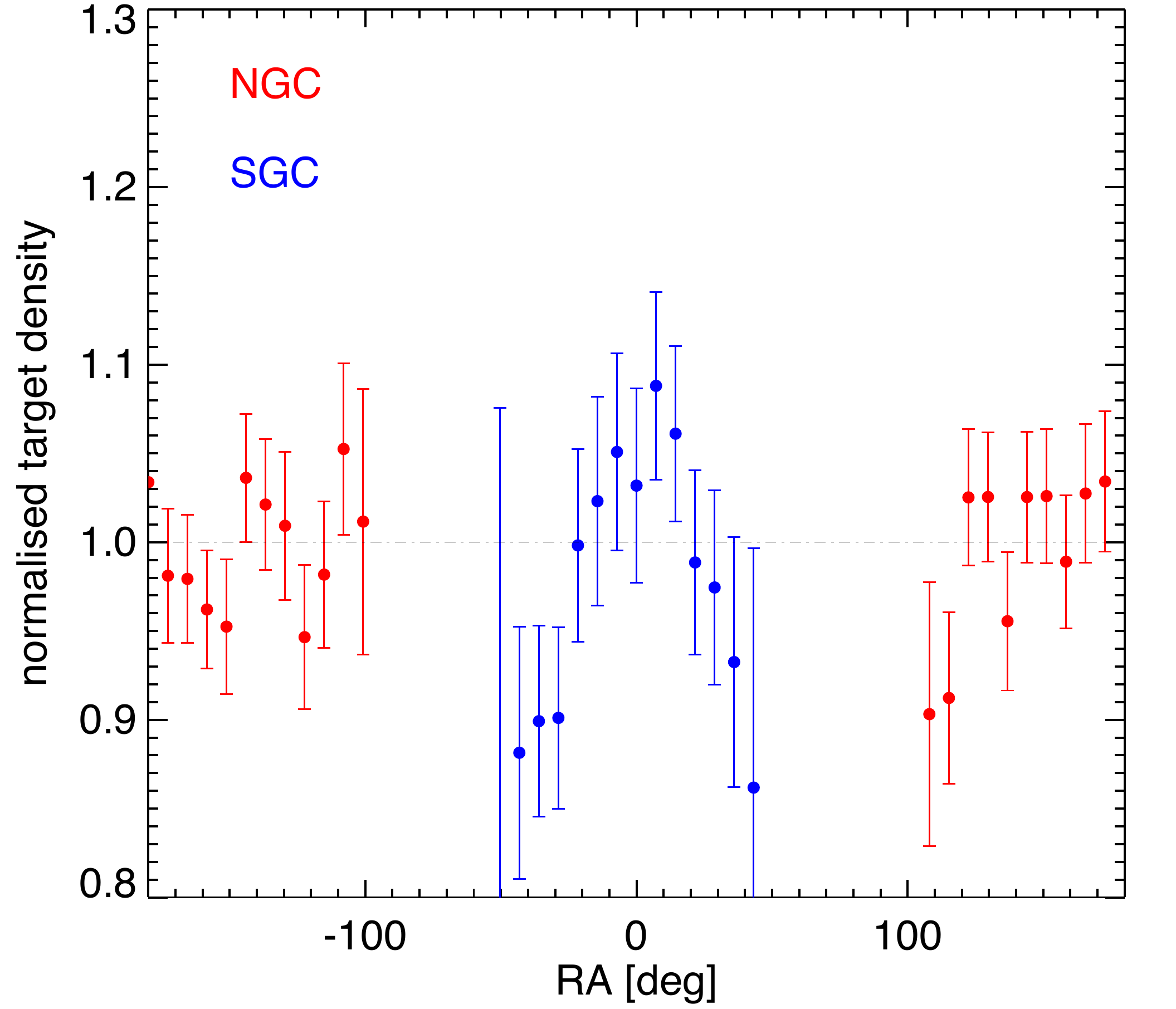}
\caption{On-sky angular target fluctuations as a function of RA for the DR11 LOWZ spectroscopic catalogue. Error bars are computed from the 1000 PTHalo mocks.} 
\label{fig:td_RA}
\end{center}
\end{figure}

While the deviation between the NGC and SGC results is consistent with the expected sample variance, we have searched for any charateristics in the LOWZ SGC data that may cause excess clustering on large scales. As mentioned in Section~\ref{sec:systematics}, weights based on stellar density, airmass, seeing or sky brightness do not significantly affect the north-south discrepancy. The only parameter that appears to correlate strongly with this excess of power is right ascension (RA) - i.e., weighting galaxies by their position in the sky along RA brings the NGC and SGC measurements into significantly better agreement. Such a weighting scheme is obviously unfavoured, as it lacks a physical motivation, but it may point to another systematic (as of yet not considered) that itself correlates with RA. Whatever the cause, we do find larger fluctuations of on-sky angular target density with RA in the SGC than we do in the NGC - see Fig.~\ref{fig:td_RA}.

According to the mocks, such fluctuations in RA are not unusual - they are compatible with a uniform target density at the 53 per-cent level in the NGC, 19 per-cent level in the SGC, and 32 per-cent over the full sky. 

A possible physical reason for these fluctuations is photometric calibration. We have shown in Section~\ref{sec:colours} that the red-sequence is well sampled and sufficiently isolated that it may be a useful standard crayon in itself - the offsets in intrinsic colours of LOWZ galaxies between the NGC and SGC are at least consistent with what is expected from \cite{SchlaflyEtAl11}. We searched for variations in intrinsic colour that would correlate with RA, and thereby affect the number density of targets through the colour cuts. Whereas we have found fluctuations with RA on the order of mmag in intrinsic $[g-r]_{0.3}$ and $[r-i]_{0.3}$, these are {\it not} significantly correlated with fluctuations in target-density. Given a lack of correlation between any physical factor and on-sky target density fluctuations, and given the fact that both the discrepancy in large-scale power between the north and south and target-density fluctuations with RA are not unusual given the mocks, we treat these fluctuations as the product of sample variance, and apply no further corrections to the data. 

Finally, we do not expect this discrepancy in large-scale power to be related to the excess of galaxies in the SGC due to photometric offsets (and discussed in Section~\ref{sec:number_density}). The increase in number density likely arises from targeting fainter galaxies, which we expect to have a lower bias. 

There are observational mechanisms that may impart subtle photometric calibration issues within the contiguous regions of sky. We plan on simulating these in the mocks for the analysis of future data releases.


\section{Analysis} \label{sec:analysis}
In this paper we concentrate on spherically-averaged two-point statistics. Anisotropic measurements are not expected to greatly improve cosmological constraints at the current signal-to-noise, though we will explore this in the final analysis of the full BOSS dataset.

\subsection{Reconstruction}

We apply the reconstruction technique (\citealt{EisensteinEtAl07}) to the data and mock catalogues in order to produce measurements that are optimised for measuring the BAO scale.
Reconstruction uses the galaxy map to construct a displacement field used to redistribute galaxies into a spatial configuration that more closely reproduces their positions had they only undergone linear growth and mitigates the effect of redshift space distortions.

The algorithm used in this paper is similar to the prescription of \cite{EisensteinEtAl07} and \cite{PadmanabhanEtAl12}.
The Lagrangian displacement field $\bold{\Psi}$ is calculated to first order using the Zel'dovich approximation applied to the smoothed galaxy overdensity field. The displacement field is corrected for redshift effects in the measured overdensity. Our implementation deviates from \cite{PadmanabhanEtAl12} slightly in that we solve for the redshift space corrected displacement field in Fourier space rather than using the finite difference method in configuration space, although we find both methods produce similar results. We refer the reader to \cite{PadmanabhanEtAl12} for more details. We use a bias value of $b=1.85$, a linear growth rate of $f=0.6413$ and a smoothing scale of $15\mpcoh$. 

The reconstruction technique has been successfully implemented using SDSS-II LRGs at $z=0.35$ by \cite{PadmanabhanEtAl12} and \cite{XuEtAl13}, who performed a spherically averaged and anisotropic BAO analysis, respectively. \cite{PadmanabhanEtAl12} achieved an improvement of a factor of 1.8 on the error on $D_V$, and \cite{XuEtAl13} reported an improvement of a factor of 1.4 on the error on $D_A$ and of 1.2 on the error on $H$, relative to the pre-reconstruction case.

\cite{Aadvark} and \cite{Abalone} successfully applied reconstruction on the DR9 CMASS sample, on an spherically averaged and anisotropic BAO analysis, respectively.  They observed only a slight reduction in the error of $D_V$, $D_A$ and $H$, when compared to the pre-reconstruction case, but at a level consistent with mock galaxy catalogues. This result can be partly explained by the fact that at higher redshift there is less to be gained as the density field is less affected by non-linearities at the BAO scale.

We present full-sky clustering measurements and covariances using both original and reconstructed catalogues. Whilst reconstruction has been shown to improve the accuracy of the BAO peak position, the effect on the full shape of two-point clustering statistics is not at the moment well studied.

 \subsection{Power-spectrum}
 
The spherically averaged power-spectrum, $P(k)$, is calculated using the \cite{FeldmanEtAl94} Fourier method, as detailed in \cite{PercivalEtAl07b} and \cite{ReidEtAl10}, and most recently implemented in \cite{Aadvark} and \cite{Aardwolf}. We maintain a box size of  $4000 \mpcoh$ and a grid that is $2048^3$, yielding an Nyquist frequency of $k\approx 1.6\hompc$, well above the maximum frequency used to fit the BAO scale (see \citealt{Aardwolf}).

As in these two most recent papers, we do not convert from galaxy density field to a halo density field, do not apply corrections for Finger-of-God effects, and do not apply luminosity-dependent weights. As such, our measurement of the power-spectrum is ideally suited for measuring the BAO scale, but care should be taken when accuracy at small scales, or modelling the full shape, is necessary.
 
The expected distribution of galaxies is modelled using the random catalogues that are constructed as described in Section~\ref{sec:randoms}. The weights applied to the random catalogue are normalised such that the total weighted number density matches for galaxies and randoms. When using full-sky catalogues (by combining NGC and SGC galaxies and randoms into single full-sky samples), this normalisation of the random weights is done independently for each Galactic cap. This approach is needed when analysing the mocks whilst using a single random catalogue (as we do) - the ratio of randoms to galaxies in each hemisphere varies slightly from one mock to the next. Grid-size, shot-noise subtraction and correction for the smoothing effect of the cloud-in-cell assignment used to locate galaxies on the grid are implemented as in \cite{Aadvark}.

The analysis of mocks post-reconstruction is performed in an identical way, with the exception that a different random catalogue is used for every mock. This procedure is necessary as post-reconstruction the random field actually contains the large-scale signal.

The spherically averaged power is computed in 1000 bins of $\Delta k = 0.0004 \hompc$, from $k=0.0002 \hompc$ to $k=0.4 \hompc$. These measurements are combined into coarser bins by taking the mean weighted by the number of $k$-modes contributing to each bin. Our fiducial binning for all analyses and plots is $\Delta k = 0.008 \hompc$, with the smallest $k$-bin centred on $k=0.008 \hompc$. This bin size was determined to be optimal in \cite{PerCov}. Our final BAO results are a weighted mean of results using $\Delta k = 0.008 \hompc$ across the different bin-centre choices.
 
 \subsection{Correlation function}
 
 We compute the two-point correlation function, $\xi(s)$, using the \cite{LandySzalay93} estimator as 
 
 \begin{equation}
 \xi(s) = \frac{DD(s) - 2DR(s) + RR(s)}{RR(s)}
 \end{equation}
 where $DD(s)$, $RR(s)$ and $DR(s)$ are normalised counts of weighted data-data, random-random and data-random pairs, separated by a distance $s$. The random catalogues are constructed as described in Section~\ref{sec:randoms}. When analysing reconstructed data and mocks, the $DR$ and $RR$ pairs in the numerator are replaced by $DS$ and $SS$, where $S$ represents the shifted particles.
 
 We compute the correlation function of the data and mocks in 200 bins linearly spaced between 1 and 200 Mpc $h^{-1}$. Without penalty we then combine our pair counts into 8 $h^{-1}$Mpc sized bins, matching the approach used in \cite{Aardwolf} and determined to be optimal for the CMASS sample in the analysis of \cite{PerCov}. This binning allows eight different choices for the bin centre. Our fiducial choice is to use a binning where the smallest scale separation is centred at $6h^{-1}$Mpc. In order to produce our final BAO measurements, we combine results from each of the eight choices.
 
\subsection{Mocks and covariance matrices}\label{sec:mocks}

We use a set of 1000 PTHalos mocks, as detailed in Manera et al. (in prep), to estimate the statistical error on our measurements.  These mocks differ from the ones used in \cite{Aardwolf} for the CMASS sample by having a halo occupation distribution (HOD) model that is allowed to vary as a function of redshift. The best-fitting HOD is found by simultaneously fitting the power and $n(z)$ of the mocks to the data. For more details see Manera et al. (in prep).



We construct covariance matrices for the power spectrum as

\begin{equation}\label{eq:sample_cov}
C_{ij} (k) = \frac{1}{N-1} \lbrack (P(k_i) - \bar{P}(k_i))(P(k_j) - \bar{P}(k_j)) \rbrack
\end{equation}

\noindent and identically for the correlation function. Error bars shown in all plots in Section~\ref{sec:results} show the square-root of the diagonal elements of these covariance matrices.

To obtain an unbiased estimate of the inverse covariance matrix $\textbf{{\sf C}}^{-1}$ we rescale
the inverse of our covariance matrix by a factor that depends on the number of
mocks and measurement bins (see e.g., \citealt{Hartlap07}) 
\begin{equation} \textbf{\sf C}^{-1} =
  \frac{N_{mocks}-N_{bins}-2}{N_{mocks}-1}~\tilde{\textbf{\sf C}}^{-1},  \label{eq:cinv}
\end{equation}

where $~\tilde{\textbf{\sf C}}^{-1}$ is the inverse of the covariance matrix in equation \ref{eq:sample_cov}.
$N_{mock}$ is 1000 in all cases, but $N_{bins}$ will change depending on the
specific test we perform. We determine $\chi^2$ statistics in the standard manner,
i.e.,  
\begin{equation}
\chi^2 = (\textbf{\it X}-\textbf{\it X}_{mod}) \textbf{\sf C}_{X}^{-1} (\textbf{\it X}-\textbf{\it X}_{mod})^{T},
\end{equation}
where the data/model vector $\textbf{{\it X}}$ can contain any combination of clustering measurements. Likelihood distributions, ${\cal L}$, are determined by assuming ${\cal L}(\textbf{\it X}) \propto e^{-\chi^2(X)/2}$. 

Building from the results of \cite{DS13}, \cite{PerCov} show that there are additional factors one must apply to uncertainties determined using a covariance matrix that is constructed from a finite number of realisations and to standard deviations determined from those realisations. Defining
\begin{equation}
A = \frac{1}{(N_{mocks}-N_{bins}-1)(N_{mocks}-N_{bins}-4)},
\end{equation}
and
\begin{equation}
B = \frac{N_{mocks}-N_{bins}-2}{A},
\end{equation}
the variance estimated from the likelihood distribution should be multiplied by
\begin{equation}
m_{\sigma} = \frac{1+B(N_{bins}-N_p)}{1+2A+B(N_p+1)},
\end{equation}
and the sample variance should be multiplied by
\begin{equation}
m_{v} = m_{\sigma}\frac{N_{mocks}-1}{N_{mocks}-N_{bin}-2}.
\end{equation}
We apply these factors, where appropriate, to all values we quote.

\section{Measuring Isotropic BAO} \label{sec:BAO_fit}
\subsection{Methodology}

We use the same methodology as \cite{Aardwolf} in order to measure the isotropic BAO position. We repeat the basic details here but refer to \cite{Aardwolf} for more detailed descriptions.

In order to measure BAO positions, we extract a dilation factor $\alpha$ by comparing our data to a template that includes the BAO and a smooth curve with considerable freedom in its shape that we marginalise over. Assuming spherical symmetry, measurements of $\alpha$ can be related to physical distances via
\begin{equation}
\alpha = \frac{D_V(z)r_d^{\rm fid}}{D^{\rm fid}_V(z)r_d}
\end{equation}
where
\begin{equation}
D_V(z) \equiv \left[cz(1+z)^2D^2_AH^{-1}\right]^{1/3},
\end{equation}
$r_d$ is the sound horizon at the baryon drag epoch, which can be accurately calculated for a given cosmology using, e.g., the software package {\sc Camb} \citep{Lewis00}, and $D_A(z)$ is the angular diameter distance. For our fiducial cosmology we find $D_{V,fid}(0.32) = 1241.47$ Mpc and $r_{d,fid} = 149.28$ Mpc.

We fit the measured, isotropically averaged, correlation function and
power spectrum separately and then combine results using the mocks to
quantify the correlation coefficient between measurements. Our fits change $\alpha$ to rescale
a model of the damped BAO in order to fit the data, while 
using polynomial terms to marginalise over the broad-band effects in
either 2-point measurement. These effects include redshift-space distortions, scale-dependent bias and any
errors made in our assumption of the model cosmology.

For both $P(k)$ and $\xi(s)$, we use the linear theory $P(k)$ produced by {\sc Camb} and split it into a smooth ``De-Wiggled'' template
$P^{\rm sm, lin}$ and a BAO template ${\rm O}^{\rm lin}$,
following \citet{EisSeoWhi07}, and using the fitting formulae of \cite{EisensteinHu98}. The damped BAO feature is then given by
\begin{equation}
{\rm O^{damp}}(k) = 1+({\rm O^{lin}}(k)-1)e^{\frac{1}{2}k\Sigma^2_{nl}}
\end{equation} 

For the $P(k)$ fits, the damping is treated as a free parameter, with a Gaussian prior of width $\pm 2h^{-1}$Mpc centred at the best-fit value recovered from the mocks. Pre-reconstruction this quantity is $\Sigma_{nl} = 8.8 h^{-1}$Mpc and post-reconstruction it is $\Sigma_{nl} = 4.6 h^{-1}$Mpc. The full model fitted for $P(k)$ is
\begin{equation}  \label{eq:mod_pk}
  P^{\rm fit}(k)=P^{\rm sm}(k){\rm O^{damp}}(k/\alpha),    
\end{equation}
where 
\begin{equation}\label{eq:Pk_poly}
    P^{\rm sm}(k)= B_p^2P(k)^{\rm sm,lin}+A_1k+A_2+\frac{A_3}{k}+\frac{A_4}{k^2}+\frac{A_5}{k^3},
\end{equation} 
These are therefore six ``nuisance'' parameters: a multiplicative
constant for an unknown large-scale bias $B_p$, and five polynomial
parameters $A_1$, $A_2$, $A_3$, $A_4$, and $A_5$. 

For the correlation function, we use a
model
\begin{equation} \label{eqn:fform}
  \xi^{\rm fit}(s) = B_\xi^2\xi^{\rm lin,damp}(\alpha s)+A_\xi(s).
\end{equation}
where $\xi^{\rm lin,damp}(s)$ is the Fourier transform of
 $P^{\rm sm, lin}(k) {\rm O^{damp}}(k)$. $B_\xi$ is a multiplicative
constant allowing for an unknown large-scale bias, while the
additive polynomial is given by
\begin{equation}
  A^\xi(s) = \frac{a_1}{s^2} + \frac{a_2}{s} + a_3
\end{equation}
where $a_i$, $1<i<3$ help marginalize over the broadband signal. Unlike for $P(k)$,
we do not allow the damping parameter to vary and instead fix it at the mean best-fit
value recovered from the mocks. In the $\xi(s)$ fits, the size of the BAO relative to $A^\xi(s)$
is allowed to vary, while in the $P(k)$ fits, the size of the BAO feature is always fixed
relative to $P^{\rm sm}$, thus the amplitude of the BAO feature has more freedom in the $\xi(s)$ independent of the damping
parameter.

The scale dilation parameter, $\alpha$, measures the relative position
of the acoustic peak in the data versus the model, thereby
characterising any observed shift.  If $\alpha>1$, the acoustic peak
is shifted towards smaller scales. For fits to both the correlation
function and power spectrum, we obtain the best-fit value of $\alpha$
assuming that $\xi(s)$ and $\log P(k)$ were drawn from multi-variate
Gaussian distributions, calculating $\chi^2$ at intervals of
$\Delta\alpha=0.001$ in the range $0.8<\alpha<1.2$. 

\subsection{Testing on Mocks}
\label{sec:mockBAO}

\begin{figure}
\resizebox{84mm}{!}{\includegraphics{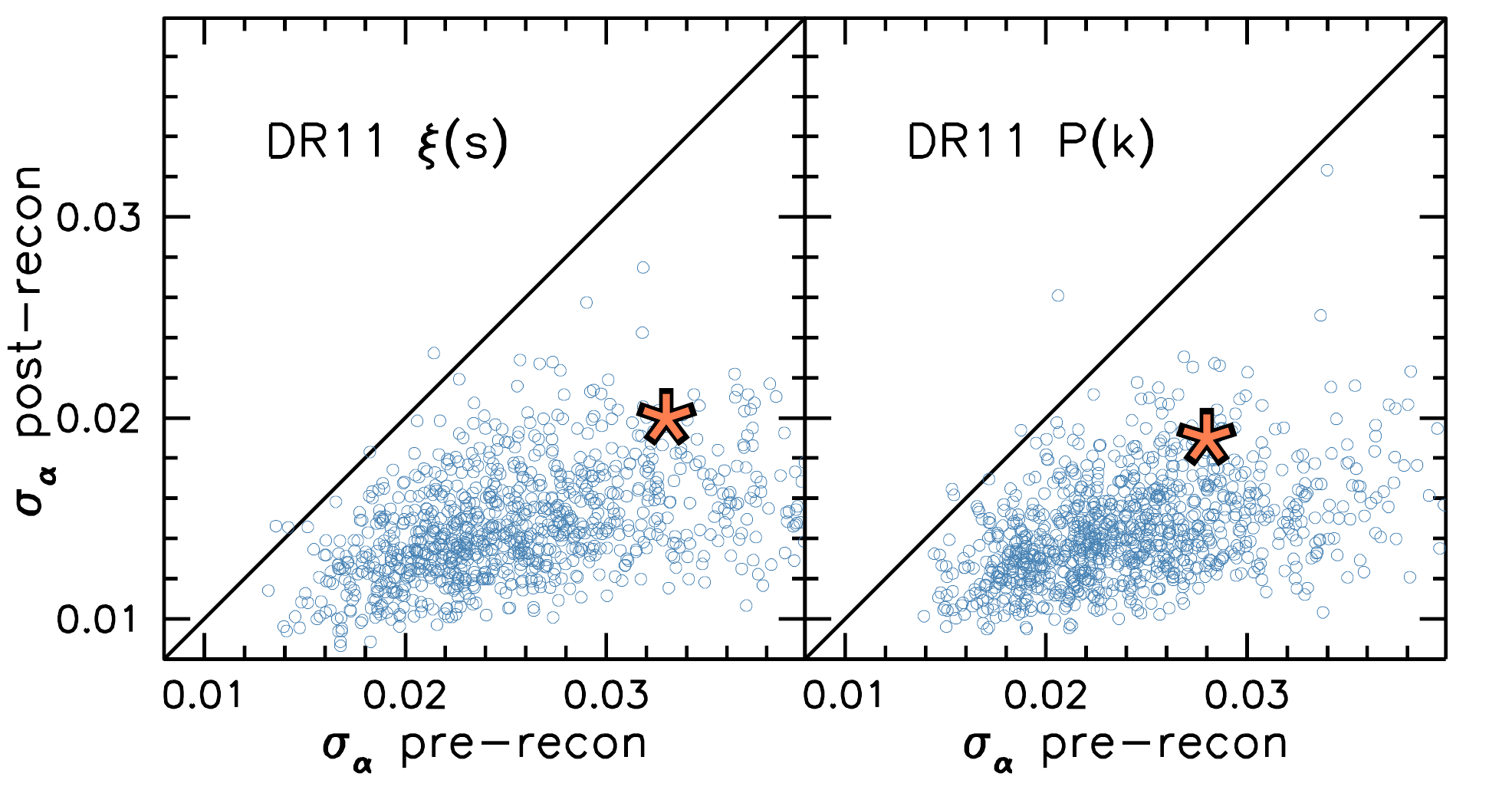}}
\resizebox{84mm}{!}{\includegraphics{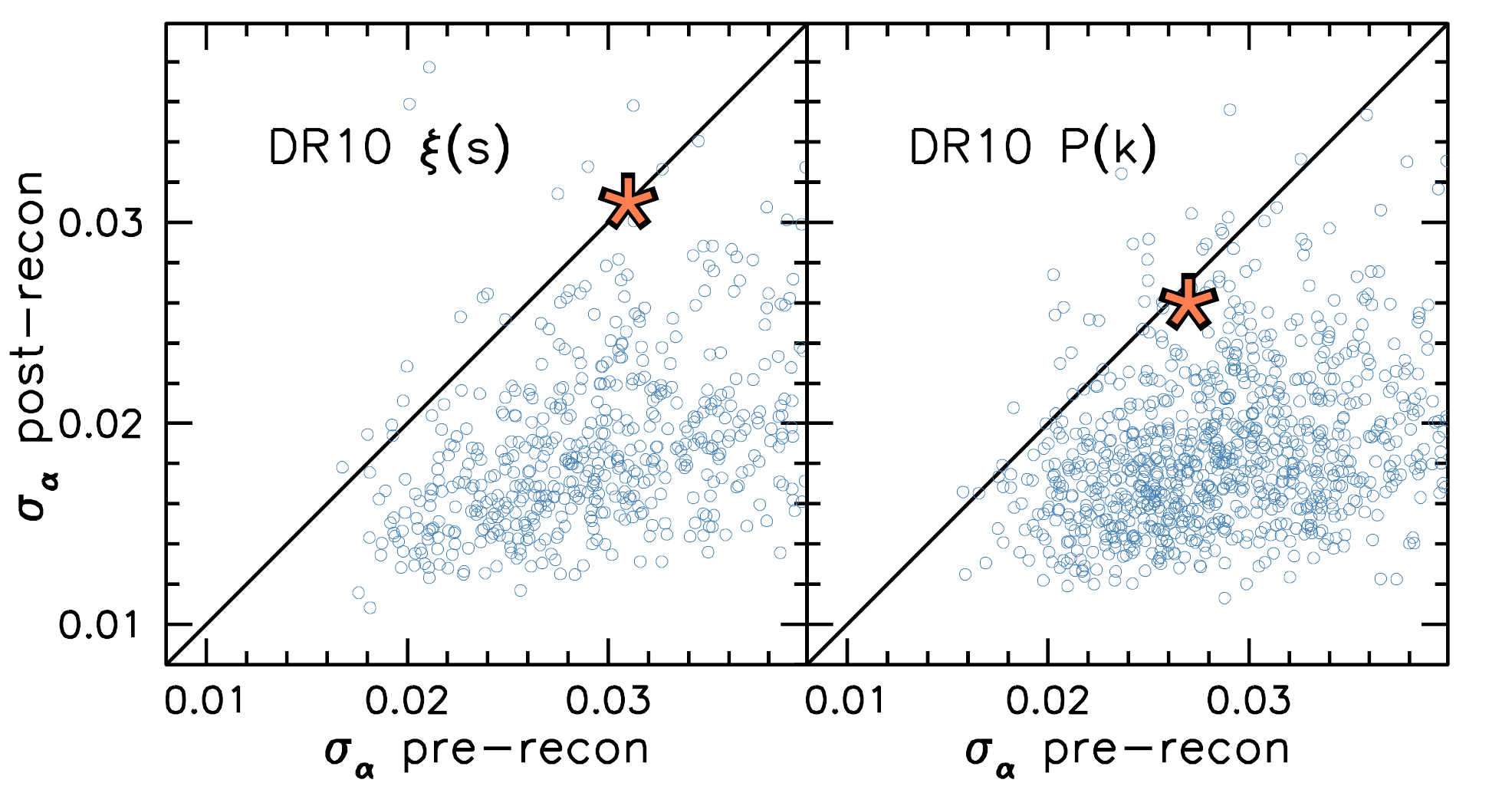}}
\caption{Plots of $\sigma_\alpha$ pre- and post-reconstruction:
mocks (circles) + data (star) for $\xi$ and $P(k)$ LOWZ DR10 and DR11. For the DR11 samples, reconstruction improves the precision in the vast majority of the 1000 mock realisations, for both $\xi(s)$ and $P(k)$, and similar improvement is recovered for the data. }
\label{fig:reconcom}
\end{figure}

\begin{table}
\caption{The statistics of isotropic BAO scale measurements recovered from the mock  galaxy samples. The parameter $\langle \alpha \rangle$ is the mean $\alpha$ value determined from 600 mock realizations of each sample, $S_{\alpha} = \sqrt{\langle(\alpha-\langle \alpha \rangle)^2\rangle}$ is the standard deviation of the $\alpha$ values, $\langle \sigma \rangle$ is the mean 1 $\sigma$ uncertainty on $\alpha$ recovered from the likelihood distribution of each realization. $S_{\sigma}$ is the standard deviation of the uncertainty recovered from the likelihood distribution. }
\begin{tabular}{lcccccc}
\hline
\hline
Estimator &  $\langle \alpha \rangle$ &$S_{\alpha}$&$\langle \sigma \rangle$&$\langle\chi^2\rangle$/dof\\
\hline
Mocks DR11 & & &  \\
Combined $P(k)$ & 0.9990 & 0.0138  & 0.0143 \\
Combined $\xi_0(s)$ & 0.9991 & 0.0135  & 0.0146 \\
post-recon $P(k)$ & 0.9989 & 0.0139  & 0.0144 & 28.0/27 \\
post-recon $\xi_0(s)$ & 0.9996 & 0.0136  & 0.0148 & 17.8/17\\
pre-recon $P(k)$ & 1.0045 & 0.0244  & 0.0241 & 27.8/27  \\
pre-recon $\xi_0(s)$ & 1.0070 & 0.0268   & 0.0266 & 18.4/17 \\
 \hline
Mocks DR10 & & & \\
post-recon $P(k)$ & 1.0004 &  0.0189 &  0.0180 & 28.0/27\\
post-recon $\xi_0(s)$ & 1.0028  & 0.0195 & 0.0180 & 17.5/17 \\
pre-recon $P(k)$ & 1.0039 & 0.0282  & 0.0298 & 27.6/27 \\
pre-recon $\xi_0(s)$ & 1.0047 & 0.0380 & 0.0374 & 15.8/17  \\
\hline
\label{tab:mockbao}
\end{tabular}
\end{table}

We test the our $\xi(s)$ and $P(k)$ isotropic BAO fitting procedure on each of our 1000 mock galaxy samples, both pre- and post-reconstruction.
The results are summarised in Table \ref{tab:mockbao}. The tests repeat those performed in \cite{Aardwolf} and we reach the same conclusion: the $P(k)$ and $\xi(s)$ BAO fits yield equally unbiased and precise BAO measurements.

Fig. \ref{fig:reconcom} displays the uncertainty obtained post-reconstruction versus that obtained pre-reconstruction, with each mock shown in a blue circle. One can see that the vast majority show improvement post-reconstruction - and more so for DR11 - and that reconstruction typically does an excellent job improving the precision of the BAO measurements obtained from mock samples. For the DR11 samples, the improvement in the mean uncertainty is 59 per cent for $P(k)$ and 85 per cent for $\xi(s)$. The improvement is dramatic for $\xi(s)$, in part due to the fact that the recovered distribution in uncertainty for the pre-reconstruction results is significantly skewed and reconstruction is particularly effective in improving the constraints on these outliers. One can further observe that the improvement afforded by reconstruction typically becomes greater with larger pre-reconstruction uncertainty.

The mean $\alpha$ that is recovered is biased low by 0.001 for $P(k)$ and 0.0008 for $\xi(s)$. The uncertainty on the ensemble mean is 0.00044, implying this bias is a 2$\sigma$ discrepancy from our expectation. Given the strong correlation between $P(k)$ and $\xi(s)$, it is unsurprising that each vary in a similar way. For the LOWZ DR10 post-reconstruction samples, the bias has an opposite sign, suggesting that the bias is not due to the modelling. The same methodology was applied to the CMASS sample \citep{Aardwolf} and no bias was found. We conclude any bias is intrinsic to the mock samples being used. The potential bias is well within the assumed systematic uncertainty of 0.003 that is added to our measurement in \cite{Aardwolf}.

As in \cite{Aardwolf}, we combine results across $\xi(s)$ and $P(k)$ measurements using different bin centres. For $\xi(s)$, these are eight bin centre choices separated by 1$h^{-1}$Mpc. For $P(k)$, these are 10 bin centre choices, each separated by 0.0008$h$Mpc$^{-1}$. The results from each bin centre are then combined based on the correlation matrix constructed from the mock samples. The process is described in detail in Section 4.3 of \cite{Aardwolf}. The measurements in each bin centre are more correlated for LOWZ than for CMASS and thus the percentage gain is smaller. The results are denoted as ``Combined'' in Table \ref{tab:mockbao}. The Combined $P(k)$ and $\xi(s)$ results have a correlation factor of 0.95, which is the same as found for CMASS in \cite{Aardwolf}.






\section{Results}\label{sec:results}
\subsection{Clustering Measurements}


\begin{figure*}
\centering
  \resizebox{0.8\textwidth}{!}{\includegraphics{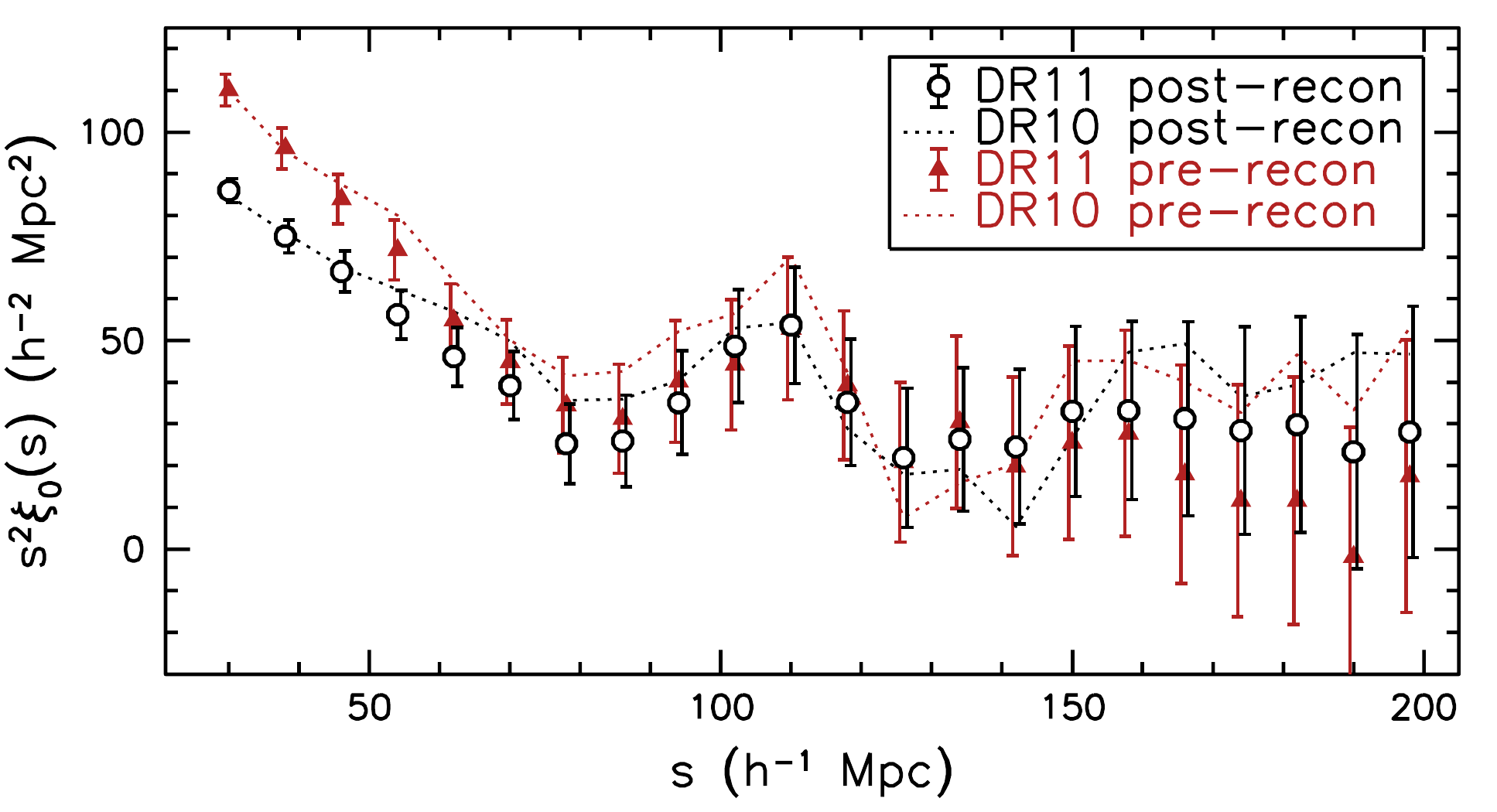}}
  \resizebox{0.8\textwidth}{!}{\includegraphics{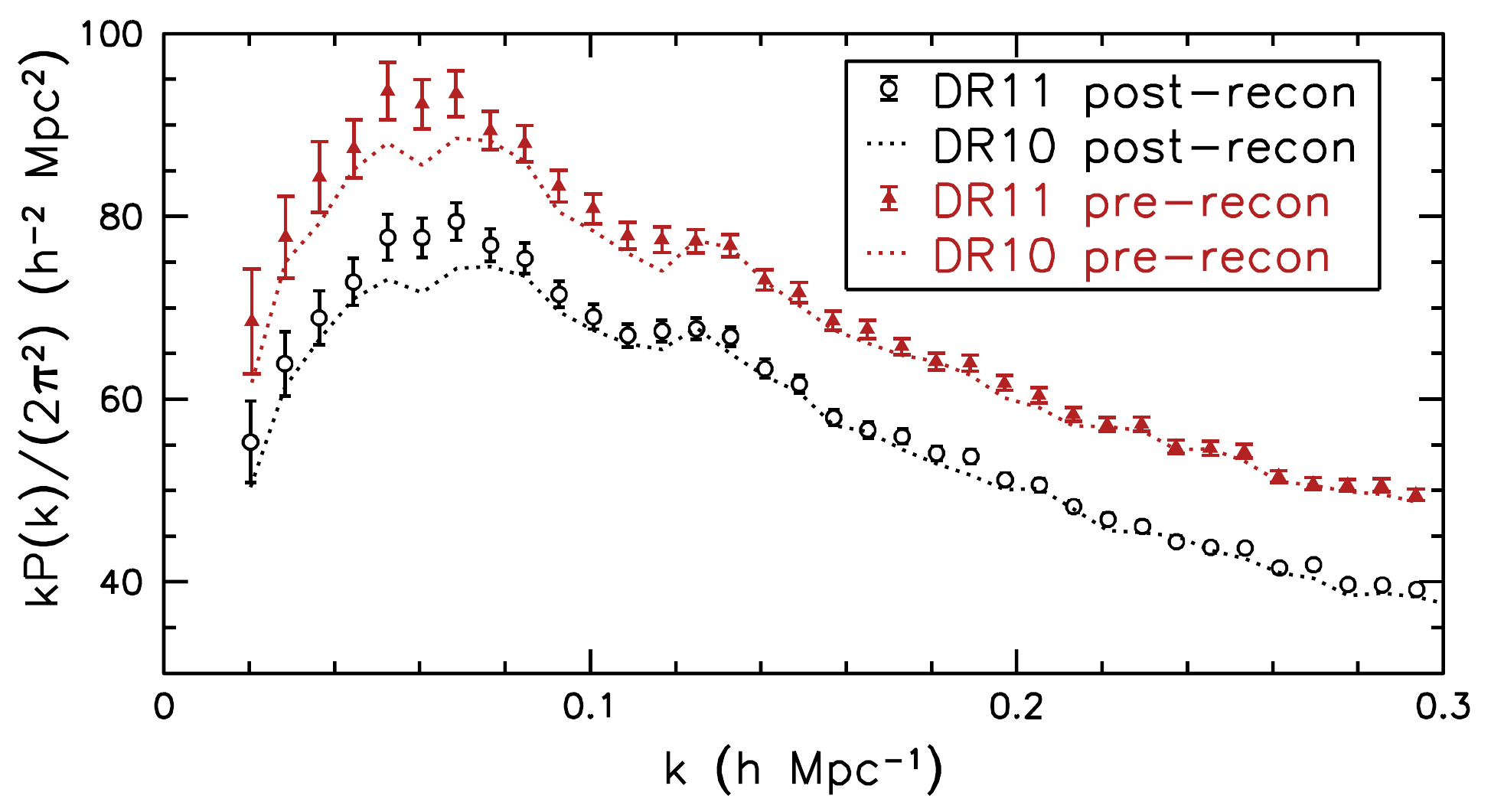}}
  \caption{Top panel: The measured monopole of the LOWZ galaxy
    correlation function, multiplied by the square of the scale, $s$,
    for BOSS data releases 10 and 11, pre- and post-reconstruction. The pre-reconstruction data
    is shown in red and the post-reconstruction data in black. The DR11 results are shown with
    points and the DR10 with dotted lines. Bottom
    panel: The measured spherically averaged LOWZ galaxy power
    spectrum, multiplied by the frequency scale, $k$, for
    BOSS data releases 10 and 11, pre- and post-reconstruction, as in the top panel.} 
\label{fig:xipkrel}
\end{figure*}

%

The configuration space and Fourier space clustering measurements made
from the DR10 and DR11 LOWZ samples are presented in
Fig.~\ref{fig:xipkrel} for $\xi(s)$ and $P(k)$, using our fiducial
binning choice and the range of scales used in our BAO fits ($30 < s < 200 h^{-1}$Mpc for $\xi(s)$ and $0.02 < k < 0.3h$Mpc$^{-1}$ for $P(k)$). The pre-reconstruction measurements are displayed using red triangles for DR11 and dotted red curves for DR10. The clustering in DR10 and DR11 are broadly consistent. In the correlation function, the power at large scales has decreased in DR11. This change is primarily due to the change in the clustering of the SGC. For the power spectrum, slightly greater power is observed in DR11 than in DR10. At the smallest $k$ (largest scales), this is partly due to the increase in the survey size, which reduces the size of the integral constraint. The post-reconstruction results are displayed using black circles for DR11 and dotted black curves for DR10. The trends are broadly consistent with those seen in the pre-reconstruction data. The amplitude of post-reconstruction is decreased at all scales for $P(k)$ and at small scales for $\xi(s)$; this is due to the removal of redshift-space distortions in the reconstructed catalog. 

\begin{figure*}
\centering
  \resizebox{0.8\textwidth}{!}{\includegraphics{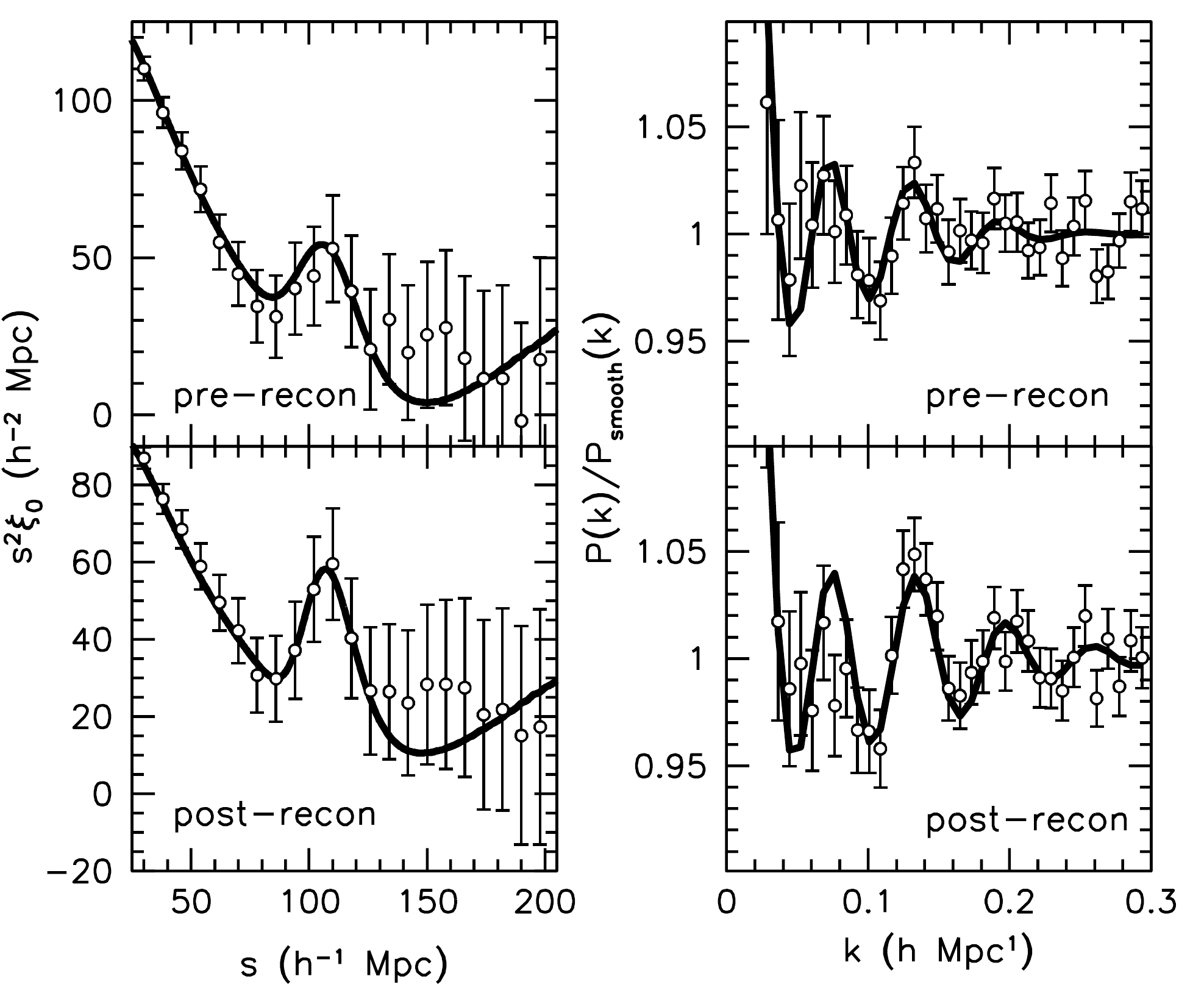}}
  \caption{ DR11 LOWZ clustering measurements (black circles) with
    $\xi(s)$ shown in the left panels and $P(k)$ in the right
    panels. The top panels show the measurements prior to
    reconstruction and the bottom panels show the measurements after
    reconstruction. The curves show the best-fit BAO model.}
\label{fig:LOWZpkxi}
\end{figure*}

\subsection{DR11 Acoustic Scale Measurement}\label{sec:BAOfits}
We find a consensus BAO measurement of $\alpha = 1.018\pm0.020$. The vast majority of mock samples show significant improvement post-reconstruction and we therefore adopt these measurements as our default. As in \cite{Aardwolf}, we measure the BAO scale by first combining the measurements of $\alpha$ from $P(k)$ and $\xi(s)$ calculated using many bin centres. These combined $P(k)$ and $\xi(s)$ measurements are then averaged to find the final consensus BAO measurement. Throughout this section, we describe the results determined at each stage, as listed in Table \ref{tab:isobaoresults}.

Fig. \ref{fig:LOWZpkxi} displays our best-fit BAO models for DR11 clustering measurement pre- (top panels) and post-reconstruction (bottom panels), using our fiducial binning. One can observe dramatic improvement of the sharpness of the BAO peak present in the post-reconstruction $\xi(s)$ measurement. For $P(k)$, the third peak becomes clear post-reconstruction, which was not noticeable pre-reconstruction. This sharpening of the BAO feature results in substantial improvement in the precision of the measurements, as reconstruction reduces the uncertainty by 47 per cent for $P(k)$ and  74 per cent for $\xi(s)$. All of the best-fit models appear to represent the data well, and this is confirmed by the fact that the minimum $\chi^2$ per degree of freedom is close to unity for each.

The BAO measurements in each of the bin centres we use are listed in Table \ref{tab:binshift}. There is little scatter in the $P(k)$ results but significant scatter in the $\xi(s)$ results. A larger scatter is expected for the $\xi(s)$ measurements, as the correlation in $\alpha$ found from separate bin centres ranges from 0.94 to 0.98 for $\xi(s)$, while the correlation ranges from 0.97 to 0.99 for the $P(k)$ alpha measurements from separate bin centres. The fit to $\alpha$ across the $\xi(s)$ bin centres has a $\chi^2$ of 11.5 for 7 degrees of freedom. A larger $\chi^2$ is expected in 12 per cent of cases, suggesting the scatter is typical.

The uncertainty on our LOWZ BAO measurement is larger than the mean result from the mocks, for both $\xi(s)$ and $P(k)$. Using the fiducial binning, we obtain $\sigma=0.019$ for both, which is 30 per cent larger than the mean uncertainty recovered from the mock samples. One can see that is result is at the edge of the distributions recovered from the mock samples by observing the orange stars in Fig. \ref{fig:reconcom}. After combining results across bin centres, the uncertainty becomes 0.021; this is the weighted mean of the uncertainties, accounting for the correlation between bin centres.
This uncertainty of 0.021 is greater than that recovered (when combining across bin centres) in all but 2.5 per cent of mock samples. The uncertainty of 0.019 found for $P(k)$ is greater than all but 5.3 per cent of mock cases.

The correlation, $C_{P,\xi}$, between the combined $\xi(s)$ and $P(k)$ results is 0.95.  Thus, the expected 1$\sigma$ dispersion in the combined $P(k)$ and $\xi(s)$ is $(\sigma^2_{\alpha,P}+\sigma^2_{\alpha,\xi}-2C_{P,\xi}\sigma_{\alpha,P}\sigma_{\alpha,\xi})^{\frac{1}{2}} = 0.007$. This value is equal to the difference we find for the combined $\xi(s)$ and $P(k)$ data, suggesting the difference is typical in magnitude. Our tests on the $P(k)$ and $\xi(s)$ measurements recovered from mock samples suggest that each are equally unbiased and precise. We therefore obtain our consensus measurement by taking the mean of the $P(k)$ and $\xi(s)$ measurements and uncertainty, yielding $\alpha = 1.018\pm0.020$. 
\begin{table}
\centering
\caption{Isotropic BAO scale measurements recovered from BOSS LOWZ data. The `Combined' results are the weighted mean of measurement across bin centres, as shown in Table~\ref{tab:binshift}. The `Consensus' results are the mean of the combined $P(k)$ and $\xi(s)$ results. }
\begin{tabular}{lccc}
\hline
\hline
Estimator  &   $\alpha$ & $\chi^2$/dof \\
\hline
DR11 & &   \\
{\bf Consensus} & {\bf $1.018\pm0.020$} & \\
Combined $P(k)$ & $1.021\pm0.019$\\
Combined $\xi(s)$ & $1.014\pm0.021$\\
post-recon $P(k)$ & $1.020\pm0.019$ & 26/27  \\
post-recon $\xi_0(s)$ & $1.012\pm0.019$ &  10/17 \\
pre-recon $P(k)$ & $1.015\pm0.028$ & 27/30   \\
pre-recon $\xi_0(s)$ & $1.016\pm0.033$ & 15/17 \\ 
\hline
DR10 & &   \\
consensus & $1.027\pm0.028$\\
post-recon $P(k)$ & $1.028\pm0.026$ &  27/27  \\
post-recon  $\xi_0(s)$ & $1.026\pm0.031$  & 13/17 \\
pre-recon $P(k)$ & $1.031\pm0.027$ & 28/30  \\
pre-recon $\xi_0(s)$ & $1.031\pm0.031$ & 22/17    \\
\hline
\label{tab:isobaoresults}
\end{tabular}
\end{table}

\begin{table}
\caption{BAO scale measurements for DR11 reconstructed data using
  different bin centres. These results are combined using their
  correlation matrix to obtain optimised BAO measurements.}
\centering
\begin{tabular}{llcc}
\hline
\hline
Shift  &   $\alpha$ & $\chi^2$/dof\\
\hline
$P(k)$ & \\
$\Delta k_i=0$ & $1.020\pm0.019$ & 26/27\\
$\Delta k_i=0.0008\hompc$ & $1.020\pm0.019$ & 27/27\\
$\Delta k_i=0.0016\hompc$ & $1.021\pm0.019$ & 26/27\\
$\Delta k_i=0.0024\hompc$ & $1.021\pm0.019$ & 24/27\\
$\Delta k_i=0.0032\hompc$ & $1.021\pm0.019$ & 24/27\\
$\Delta k_i=0.004\hompc$ & $1.021\pm 0.019$ & 20/27\\
$\Delta k_i=0.0048\hompc$& $1.021\pm0.019$ & 20/27\\
$\Delta k_i=0.0056\hompc$ & $1.021\pm 0.019$ & 21/27\\
$\Delta k_i=0.0064\hompc$ & $1.020\pm0.019$ & 23/27\\
$\Delta k_i=0.0072\hompc$ & $1.021\pm 0.019$ & 26/27\\
\hline
 $\xi(s)$ &  \\
  $\Delta s_i=-2\mpcoh$ & $1.019\pm0.019$ & 10/17\\
$\Delta s_i=-1\mpcoh$ & $1.014\pm0.018$ & 10/17\\
$\Delta s_i=0$  &  $1.012\pm0.019$ & 10/17  \\
$\Delta s_i =+1\mpcoh$ & $1.004\pm0.020$ & 16/17\\
 $\Delta s_i=+2\mpcoh$ & $1.006\pm0.024$ & 22/17\\
$\Delta s_i=+3\mpcoh$ & $1.016\pm0.026$& 21/17\\
$\Delta s_i=+4\mpcoh$ & $1.018\pm0.022$ & 19/17\\
$\Delta s_i=+5\mpcoh$ & $1.024\pm0.020$ & 15/17\\
\hline
\label{tab:binshift}
\end{tabular}
\end{table}

\subsection{DR10 Acoustic Scale Measurement}

For completeness, we also include DR10 BAO measurements, determined using our fiducial binning. The results are listed in Table \ref{tab:isobaoresults}. Taking the mean of the $P(k)$ and $\xi(s)$ results, we find the consensus result for DR10 of $\alpha = 1.027\pm0.028$.

As for DR11, both the post-reconstruction $P(k)$ and $\xi(s)$ yield results that are significantly worse than the mean results from the mocks, as can be seen by observing the orange stars in the bottom panels of Fig. \ref{fig:reconcom}. We find 5.1 per cent of the DR10 mocks yield an uncertainty that is greater 0.031 for $\xi(s)$ and 6.2 per cent yield an uncertainty that is greater than 0.026 when using the $P(k)$ measurements. Pre-reconstruction, the uncertainty on the DR10 measurements are each slightly better than the mean recovered from the mocks. Unlike for DR11, reconstruction does not significantly improve the LOWZ DR10 BAO measurements. The DR11 area is more contiguous, and one can see in Fig. \ref{fig:reconcom} that the precision improves for a larger fraction of mock samples post-reconstruction for DR11 than DR10.

The DR10 measurement is consistent with the DR11 measurement. LOWZ DR10 covers 70 per cent of the LOWZ DR11 footprint. Assuming a correlation 0.7, the 0.009 difference is well within the expected 1$\sigma$ variation between LOWZ BAO measurements from the two data releases of 0.020. 
\begin{table}
\caption{Robustness checks on Isotropic BAO scale measurements
  recovered from DR11 reconstructed data. }
\centering
\begin{tabular}{llcc}
\hline
\hline
Estimator & Change  &   $\alpha$ & $\chi^2$/dof\\
\hline

$P(k)$ & fiducial & $1.020\pm0.019$ & 25.9/27\\
&NGC only & $1.037\pm0.022$ & 29.3/27\\
&SGC only & $0.988\pm0.033$ & 29.1/27\\
& $0.02<k<0.25\hompc$ & $1.019\pm0.019$ & 13.6/21\\
& $0.02<k<0.2\hompc$ & $1.016\pm0.019$ & 8.9/15\\
& $0.05<k<0.3\hompc$ & $1.024\pm0.018$ & 22.8/23\\
& $\Sigma_{nl} = 3.8\pm0.0$ & $1.020\pm0.016$ & 27.7/28\\  
& $\Sigma_{nl} = 5.8\pm0.0$ & $1.020\pm0.018$ & 25.6/28\\  
& $A_1, A_2 = 0$ & $1.025\pm0.019$ & 30.3/29\\
& Spline fit & $1.022\pm0.018$ & 24.4/24\\
& $\Delta k = 0.0032\hompc$ & $1.017\pm0.019$ & 83.3/79\\
 & $\Delta k = 0.004\hompc$ & $1.024\pm0.019$ & 62.4/62\\
 & $\Delta k = 0.006\hompc$ & $1.020\pm0.019$ & 32.0/39\\
 & $\Delta k = 0.01\hompc$ & $1.017\pm0.019$ & 9.9/20\\
 & $\Delta k = 0.012\hompc$ & $1.019\pm0.020$ & 7.3/15\\
 & $\Delta k = 0.016\hompc$ & $1.014\pm0.019$ & 3.6/9\\
 & $\Delta k = 0.02\hompc$ & $1.024\pm0.022$ & 2.0/6\\
 \hline
$\xi(s)$ & fiducial & $1.012\pm0.019$ & 9.9/17\\
&NGC only & $1.031\pm0.026$ & 14/17\\
&SGC only & $0.989\pm0.028$ & 15/17\\ 
 &  $50 < s < 150\mpcoh$ & $1.012\pm0.018$ & 7/7\\
 & $a_1, a_2, a_3 = 0$ & $1.005\pm0.018$ & 16.7/20\\
 & $a_1, a_2 = 0$ & $1.009\pm0.018$ & 10.5/19\\
 & $a_1 = 0$ & $1.011\pm0.018$ & 9.9/18\\
 & $a_2 = 0$ & $1.012\pm0.019$ & 10.0/18\\
 & $B_\xi$ free & $1.011\pm0.019$ & 9.9/17\\
 & $\Sigma_{nl} = 3.8\mpcoh$ & $1.010\pm0.019$ & 10.2/17\\
 & $\Sigma_{nl} = 5.8\mpcoh$ & $1.013\pm0.019$ & 9.9/17\\
 &  $\Delta s = 4\mpcoh$ & $1.013\pm0.022$ & 51/38\\
&  $\Delta s = 5\mpcoh$ & $1.014\pm0.023$ & 29/29\\
&  $\Delta s = 6\mpcoh$ & $1.011\pm0.020$ & 38/23\\
&  $\Delta s = 7\mpcoh$ & $1.014\pm0.020$ & 17/19\\
&  $\Delta s = 9\mpcoh$ & $0.998\pm0.022$ & 10/14\\
&  $\Delta s = 10\mpcoh$ & $1.016\pm0.027$ & 9/12 \\
\hline
\label{tab:isorobust}
\end{tabular}
\end{table}

\subsection{Robustness Tests}\label{sec:baorobust}

We subject our DR11 post-reconstruction to a variety of robustness tests, listed in Table \ref{tab:isorobust}. We measure the BAO scale in the NGC and SGC separately. For the $\xi(s)$ fiducial bin choice, we find $\alpha = 1.031\pm0.026$ in the NGC and $\alpha = 0.989\pm0.028$ in the SGC. The discrepancy in the two measurements is only 1.1$\sigma$. The difference has the opposite sign of the difference found between the DR11 CMASS post-reconstruction NGC and SGC BAO scale measurements in \cite{Aardwolf}. The difference in the NGC and SGC BAO measurements found using $P(k)$ measurements is slightly larger and represents a 1.2$\sigma$ offset.

Only minor changes in the BAO scale are measured using $\xi(s)$ when we change the fitting details. The measurement does not change when we reduce the fit range to $50 < s < 150h^{-1}$Mpc, suggesting that, as expected, the BAO information lies entirely within this range. The BAO measurement changes by only 0.001 ($0.05\sigma$) when we allow the term that modulates the amplitude of the BAO template, $B_{\xi}$, to vary freely. Similarly small changes are found when we change the damping parameter $\Sigma_{nl}$ by $\pm1h^{-1}$Mpc.

The power spectrum results change by an insignificant degree when the range in $k$ used for the fits is changed, as all variations are less than $0.23\sigma$. No change in $\alpha$ is seen when $\Sigma_{nl}$ is fixed, but we do find this reduces the recovered uncertainty. This result is expected, especially when $\Sigma_{nl}$ is set to a smaller value than the best-fit for mocks, as this template will have a more pronounced BAO feature and may artificially reduce the allowed range in $\alpha$. Indeed, it was found in \cite{Aadvark} that a marginalization over $\Sigma_{nl}$ was required for the $P(k)$ fits in order to recover a Gaussian distribution of $(\alpha-1)/\sigma_{\alpha}$ for results recovered from the mock realisations.

The number of terms used in the polynomial that allows marginalization over the broadband shape has a small affect on the recovered BAO scale. The data show a strong preference for including a constant term in the fit, as the minimum $\chi^2$ decreases by 6.2 comparing the case where no polynomial is used ($a_1,a_2,a_3 = 0$) to the case where the constant term, $a_3$, is included. The measurement also increases by $0.25\sigma$ moving between the two cases. Adding a second term to the polynomial (either $a_1$ or $a_2$) reduces the $\chi^2$ by only 0.5 and increases $\alpha$ by at most 0.003. Negligible changes are found when the three term polynomial is used, compared to using either of the two term polynomials, suggesting the measurement is stable once the order of the polynomial used in the fit is at least two.

We find a negligible change ($0.1\sigma$) in the recovered BAO scale when we apply the `spline fit' technique applied in \cite{Aadvark}, who use a spline fit to the smooth component of the $P(k)$ model instead of a polynomial. A small increase ($0.26\sigma$) is produced in the measured BAO scale when we set the polynomial terms $A_1$ and $A_2$ to zero. It was found in \cite{Aardwolf} that the $A_1$ and $A_2$ terms were necessary to provide a good fit to the smooth component of mock $P(k)$, and thus such a small change in the recovered value of $\alpha$ is not of serious concern. The data show a small preference for the inclusion of $A_1$ and $A_2$ in the model, as the $\chi^2$ is reduced by 4.4 in this (fiducial) case.

The $\xi(s)$ results show some scatter when the bin size used is changed. The changes in $\alpha$ from the fiducial measurement are all $\leqslant 0.004$, except for the measurement using $9h^{-1}$Mpc bins, which decreases $\alpha$ by 0.014. The mean result averaged across all of the bin size choices is $\alpha = 1.011\pm0.022$, which is close to our weighted mean measurement across the $\xi(s)$ bin centre choices.

The $P(k)$ BAO measurements vary by as much as $0.032\sigma$ from the fiducial measurement when one changes the bin sizes of the $P(k)$ measurement. This level of variation is consistent with the correlation found between the results recovered from mock $P(k)$ measurements, as $0.32\sigma$ fluctuations are expected for measurements with a 0.95 correlation. None of our robustness tests find significant variations in the recovered BAO scale measurements. We conclude that for both $P(k)$ and $\xi(s)$, our LOWZ DR11 BAO measurements are robust to reasonable changes in the methodology used to obtain the measurement.

\subsection{Comparison with previous results}

\cite{PadmanabhanEtAl12} measured the acoustic scale at $z=0.35$ using the correlation function of SDSS-II LRGs. Pre-reconstruction they achieved a 3.5\% measurement of $D_V$, and 1.9\% post-reconstruction. Although the total area used is similar to that used here, the LRG measurement is shot-noise dominated due to the relatively low number density of the sample. We therefore expect that in the pre-reconstruction case the LOWZ sample will produce a more accurate result, which is indeed the case. The improvement obtained from reconstructing the density-field is, however, larger in the case of the DR7 LRGs. This change is likely due to the fact that the DR7 area is more contiguous than the DR11 area used here. The two values are in agreement with one another: \cite{PadmanabhanEtAl12} measure $D_V(0.35)/r_d = 8.88\pm 0.17$; if we adjust this to $z=0.32$ using the best-fit $\Lambda$CDM model, $\alpha = 1.012 \pm 0.019$, close to the DR11 value of $\alpha = 1.018 \pm 0.020$ found here. There is no overlap in redshift between the LOWZ and CMASS samples, unlike between the DR7 LRG and the BOSS CMASS samples, allowing us to treat the samples as being independent. This, combined with the lower mean redshift of the LOWZ sample, means that our measurement provides the most constraining power when combined with the CMASS measurements at $z=0.57$.

\section{Summary and conclusions}\label{sec:conclusions}
In this paper we present the first large-scale clustering measurements for the LOWZ sample of the SDSS-III BOSS survey, and have measured the acoustic scale at $z=0.32$. Our findings are:

\begin{itemize}
\item There are no significant trends of target-density with stellar density or other parameters. Unlike CMASS, we currently do not apply systematic weights to the LOWZ galaxies.
\item There are differences between the NGC and the SGC in terms of number densities and large-scale clustering power. The former is likely related to offsets in photometric calibration between the two hemispheres. We treat the two samples independently to overcome this issue, and we furthermore demonstrate that  despite these offsets there is no significant difference between the type of galaxies targeted. The differences in the large-scale power measured in the two regions are consistent with the sample variance estimated from the mocks.
\item The excess of power in the SGC, in spite of being consistent with cosmic variance, is correlated with target fluctuations as a function of Right Ascension. These fluctuations are not unusual given the mocks, but they will be studied in further detail in the final data release.
\item The fits of the BAO scale are consistent between the correlation and power-spectrum, and both estimators are unbiased and robust to changes in bin size, scales used for the fit, damping parameter, and details of the model template. The BAO scales measured in each Galactic cap are also consistent with one another.
\item Reconstruction gives an improvement on our final error on the DR11 BAO scale of over 70\% for the correlation function and 50\% for the power-spectrum. The improvement is much better than in the DR10 measurements, due an improvement on the survey footprint.
\item When combined with the CMASS measurements presented in \cite{Aardwolf}, the final measurement of distance scale presented here, $D_V(0.32)= 1.264 \pm 25 (r_d/r_{d,fid})$, is the most constraining at these intermediate redshifts.
\end{itemize}

\section*{Acknowledgements}
RT is thankful for support from the European Research Council and the Science \& Technology Facilities Council. Numerical computations were done on the Sciama High Performance Compute (HPC) cluster which is supported by 
the ICG, SEPNet and the University of Portsmouth. 

Funding for SDSS-III has been provided by the Alfred P. Sloan Foundation, the Participating Institutions, the National Science Foundation, and the U.S. Department of Energy Office of Science. The SDSS-III web site is http://www.sdss3.org/.

SDSS-III is managed by the Astrophysical Research Consortium for the Participating Institutions of the SDSS-III Collaboration including the University of Arizona, the Brazilian Participation Group, Brookhaven National Laboratory, Carnegie Mellon University, University of Florida, the French Participation Group, the German Participation Group, Harvard University, the Instituto de Astrofisica de Canarias, the Michigan State/Notre Dame/JINA Participation Group, Johns Hopkins University, Lawrence Berkeley National Laboratory, Max Planck Institute for Astrophysics, Max Planck Institute for Extraterrestrial Physics, New Mexico State University, New York University, Ohio State University, Pennsylvania State University, University of Portsmouth, Princeton University, the Spanish Participation Group, University of Tokyo, University of Utah, Vanderbilt University, University of Virginia, University of Washington, and Yale University.

\bibliographystyle{mn2e}
\bibliography{my_bibliography}

\end{document}